\documentclass[british,prb,twocolumn]{revtex4}
\usepackage[latin9]{inputenc}
\usepackage{color}
\usepackage{babel}
\usepackage{amsmath}
\usepackage{amssymb}
\usepackage{graphicx}
\usepackage[unicode=true,pdfusetitle,bookmarks=true,bookmarksnumbered=false,bookmarksopen=false,
breaklinks=false,pdfborder={0 0 1},backref=false,colorlinks=false]{hyperref}

\makeatletter
\@ifundefined{textcolor}{}
{%
 \definecolor{BLACK}{gray}{0}
 \definecolor{WHITE}{gray}{1}
 \definecolor{RED}{rgb}{1,0,0}
 \definecolor{GREEN}{rgb}{0,1,0}
 \definecolor{BLUE}{rgb}{0,0,1}
 \definecolor{CYAN}{cmyk}{1,0,0,0}
 \definecolor{MAGENTA}{cmyk}{0,1,0,0}
 \definecolor{YELLOW}{cmyk}{0,0,1,0}
}

\makeatother

\begin{document}

\title{Delayed entanglement echo for individual control of a large number
of nuclear spins}

\author{Zhen-Yu Wang, Jorge Casanova, and Martin B. Plenio}

\affiliation{Institut f\"ur Theoretische Physik and IQST, Albert-Einstein-Allee
11, Universit\"at Ulm, D-89069 Ulm, Germany }

\maketitle
\textbf{
Methods for achieving quantum control and detection of individual
nuclear spins by single electrons of solid-state defects play a central 
role for quantum information processing and nano-scale nuclear magnetic 
resonance (NMR)~\cite{maurer2012room,zhao2012sensing,kolkowitz2012sensing,taminiau2012detection,liu2013noise,taminiau2014universal,waldherr2014quantum,london2013sense}.
However, with standard techniques, no more than 8 nuclear spins have 
been resolved~\cite{zhao2012sensing,kolkowitz2012sensing,taminiau2012detection,schirhagl2014nitrogen,laraoui2013high}.
Here we develop a new method that improves significantly the ability
to spectrally resolve nuclear spins and demonstrate its capabilities 
with detailed numerical simulations by using a nitrogen-vacancy (NV) 
centre~\cite{doherty2013nitrogen} as model system. Based on delayed 
entanglement control, a technique combining microwave and radio-frequency 
(rf) fields, nuclei with resonances in a broad frequency band can be 
unambiguously~\cite{loretz2015spurious} and individually addressed by 
the sensor electron. Additionally the spectral resolution can extend 
beyond the electron spin relaxation time by using a long-lived qubit 
memory. Our method greatly increases the number of useful register 
qubits accessible to a defect centre and improves the signals of 
nano-scale NMR.} 

Nuclear spins are natural quantum bits with long coherence times for quantum
information tasks~\cite{zhong2015optically} and they encode information about
the structure of molecules and materials in a form that is accessible to NMR 
techniques~\cite{mehring2012principles}. The NV centre in diamond represents 
a promising nano-scale platform for detection and coherent control of such 
nuclear spins~\cite{doherty2013nitrogen,schirhagl2014nitrogen}. In type
IIa diamonds, the decoherence of the NV electron spin is dominated by the
presence of $^{13}\text{C}$ nuclei. However, when properly controlled, the
$^{13}\text{C}$ nuclear spins in the vicinity of an NV centre become useful 
resources~\cite{liu2013noise,taminiau2014universal,waldherr2014quantum}.
Furthermore, NV centres can be implanted close to the diamond surface
to detect the signal of nuclear spins above the surface~\cite{muller2014nuclear,lovchinsky2016nuclear},
which opens opportunities for both quantum information processing
~\cite{taminiau2014universal,waldherr2014quantum,cai2013large} 
and single molecule NMR~\cite{lovchinsky2016nuclear} when environmental noise 
can be controlled. 

Originally developed in NMR, dynamical decoupling (DD) techniques 
~\cite{mehring2012principles,yang2011preserving} 
can extend significantly qubit coherence times and they can also be applied to
address single nuclear spins by an NV centre~\cite{kolkowitz2012sensing,taminiau2012detection,zhao2012sensing,london2013sense,cai2013nmr,casanova2015robust,wang2015positioning}.
Nevertheless standard DD techniques can only be used to address a few
nuclear spins because of a number of drawbacks such as low spectral resolution~\cite{laraoui2013high},
resonance ambiguities~\cite{loretz2015spurious}, and perturbations
from the electron-nuclear coupling~\cite{casanova2015robust}. In this
respect correlation spectroscopy could improve the resolution by measuring
the NV signal over long evolution times~\cite{laraoui2013high}, however
this technique does not provide advantages on individual spin addressing 
and control.

Our method overcomes these difficulties by selectively addressing
target nuclear spins by radio-frequency (rf) fields in a delay window
while the entanglement with the electron spin sensor is preserved
by a subsequent Hahn echo~\cite{mehring2012principles} operation. In this
manner highly selective entangling gates between the electron spin
and different target nuclear spins can be achieved.

\begin{figure*}
\includegraphics[clip,width=0.9\textwidth]{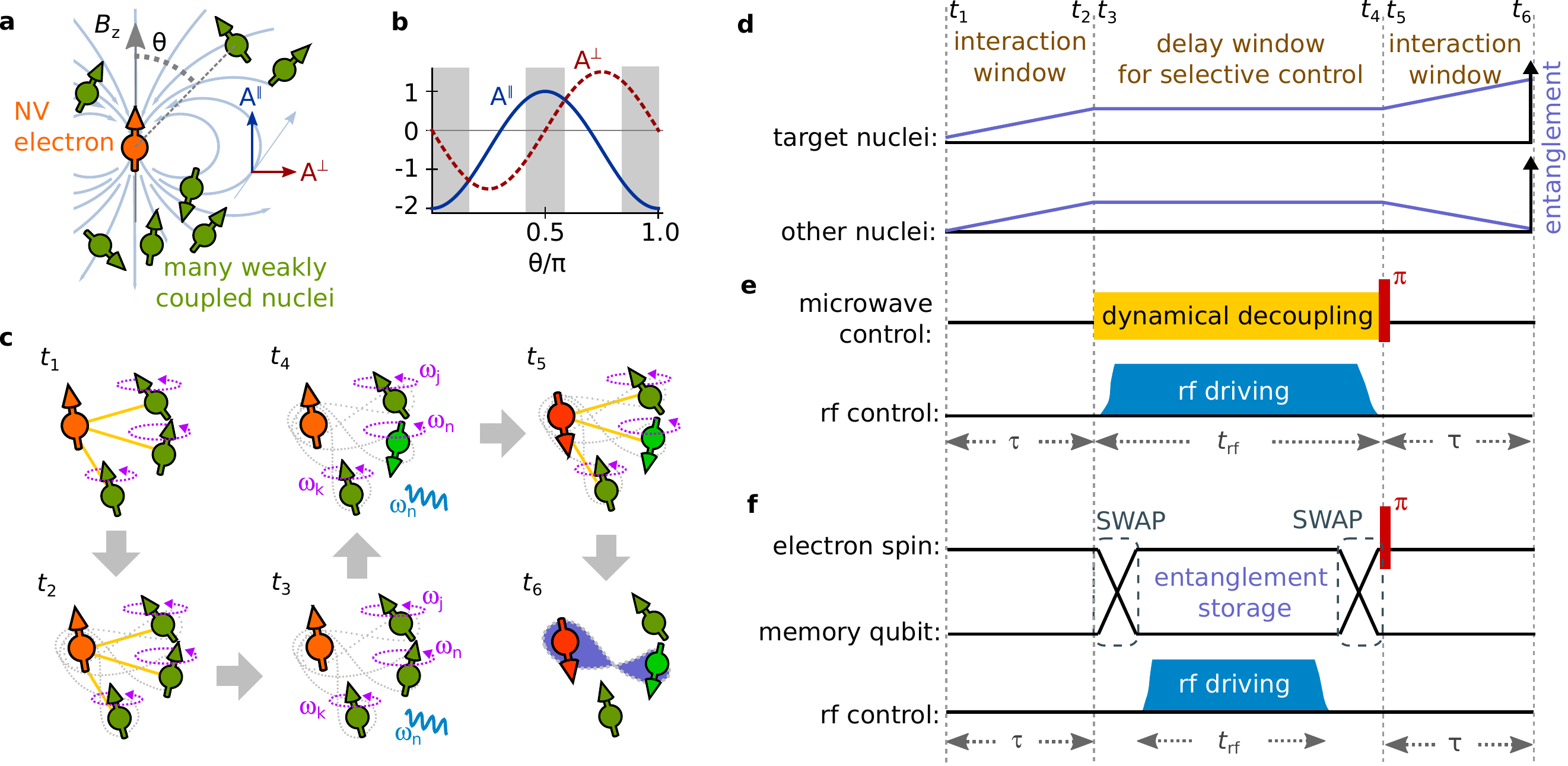}\caption{\label{fig:FigSketch} 
\textbf{Delayed entanglement control between
the NV electron and its surrounding nuclear spins.} \textbf{a}, A
large number of nuclear spins weakly coupled to the electron spin
of NV centre by the hyperfine field. \textbf{b}, Relative amplitudes
of the components of the hyperfine field. The perpendicular component
$A^{\perp}$ is weaker than the parallel component $A^{\parallel}$
at the shaded regions of the plot. \textbf{c}, Illustration of the
spin states in the process of delayed entanglement echo, with the
entanglement changes sketched in \textbf{d}. Implementation of the
delay window by DD (\textbf{e}) or by storing the entanglement to
a long-lived memory qubit (\textbf{f}).}
\end{figure*}

\begin{figure*}
\includegraphics[clip,width=0.9\textwidth]{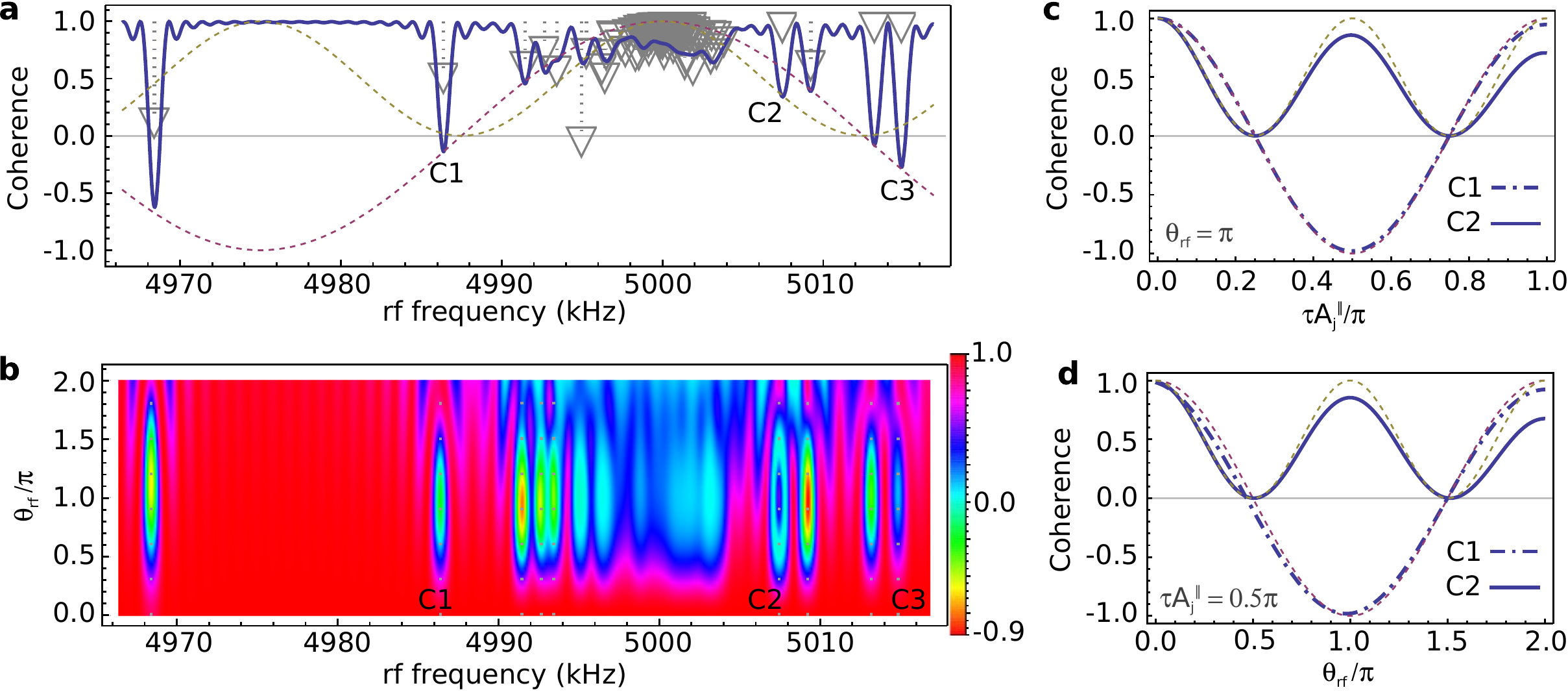}\caption{\label{fig:FigSpectrum} 
\textbf{Coherence signals of delayed entanglement echo by DD.} \textbf{a}, Coherence dips 
of NV electron qubit by using the electron spin levels $m_{s}=0,1$ for the interaction 
time $\tau=10$ $\mu$s and DD protected rf driving with single-spin rotation angle 
$\theta_{\text{rf}}=\pi$. The dashed lines shows the coherence signal $L^{p}(\omega_{\text{rf}} 
-\omega_{^{13}\text{C}},\theta_{\text{rf}})$ for one ($p=1$) and two ($p=2$) spins. 
\textbf{b}, As in \textbf{a} but changing $\theta_{\text{rf}}$ and transferring the NV 
electron qubit to the levels $m_{s}=\pm1$ for the interaction windows with $\tau=13$ $\mu$s.
\textbf{ c},\textbf{d}, Coherence oscillations when changing $\tau$ and $\theta_{\text{rf}}$
using the scheme in \textbf{b}. The arrows (with the length propositional to $A_{j}^{\perp}$) 
in \textbf{a} and the vertical dotted lines in \textbf{b} show the precession frequencies 
$\omega_{j}$ (bare Larmor frequency $\omega_{^{13}\text{C}}=2\pi\times5$ MHz). }
\end{figure*}

\begin{figure*}
\includegraphics[clip,width=0.9\textwidth]{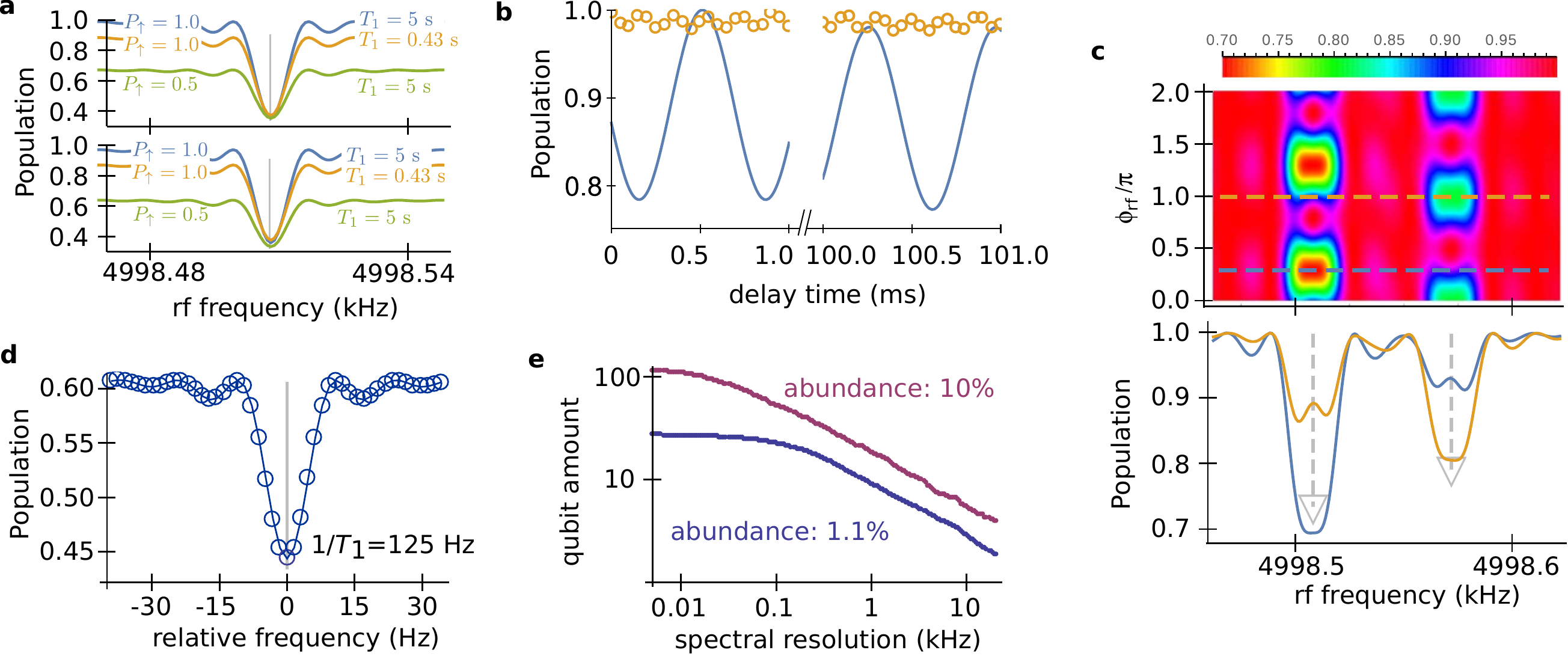}\caption{\label{fig:FigAncilla}\textbf{Delayed entanglement echo by using
a nuclear memory qubit.} \textbf{a}, Signal dips when addressing to
a single distant $^{13}\text{C}$ nuclear spin $\approx2.2$ nm away
from the NV centre by storing the qubit state in a $^{13}\text{C}$
(upper panel) or in the intrinsic $^{14}\text{N}$ (lower panel) memory
qubit during the delay windows, for $t_{\text{rf}}=100$ ms, $\tau=100$
$\mu$s, and various low-temperature electron spin relaxation times
$T_{1}$ and initial populations $P_{\uparrow}$ of the memory qubit
state $|\uparrow\rangle$. \textbf{b}, Population modulation (solid
line) caused by two interacting $^{13}\text{C}$ nuclei in a C-C bond
$\sim1.5$ nm away from the NV centre with $\tau=200$ $\mu$s and
a $^{14}\text{N}$ memory. Application of LG decoupling suppresses
nuclear dipolar interactions and hence the modulation (circles). \textbf{c},
Signal patterns from two uncoupled distant $^{13}\text{C}$ spins
($2.2$ and $2.3$ nm away from the NV centre) when using a $^{14}\text{N}$
memory qubit and DD to protect the interaction windows of duration
$\tau=500$ $\mu$s (see Supplementary Information for the method).
Lower panel shows the cross-sectional plots of upper panel (dashed
lines) when the initial rf phases $\phi_{\text{rf}}$ match the azimuthal
angles of addressed nuclear spins. \textbf{d}, Room-temperature signal
of one proton spin placed 4 nm away from the NV centre using a $^{13}\text{C}$
memory qubit and optical illumination during the delay window (see
Supplementary Information). With the rf driving of $t_{\text{rf}}\approx10T_{1}$,
the line-width of the signal is well beyond the limit set by electron
spin relaxation time $1/T_{1}$. One major reason of the reduction
of signal contrast is the leakage out from the electron qubit to another
electron spin triplet level. \textbf{e}, Log-log plot of the average number of individual
$^{13}\text{C}$ register qubits with $A_{j}^{\parallel}>4$ kHz and
$A_{j}^{\parallel}$ different from other spins by amounts larger
than the spectral resolution, by averaging over 1000 samples. A magnetic
field is used for the bare Larmor frequency $\omega_{^{13}\text{C}}=2\pi\times5$
MHz. In \textbf{a},\textbf{d},\textbf{c}, $\theta_{\text{rf}}=\pi$,
and $\theta_{\text{rf}}=0$ in \textbf{b}. In \textbf{a},\textbf{b},
the electron qubit state $|\downarrow_{e}\rangle=|0\rangle$ is transferred
to $|\downarrow_{e}\rangle=|-1\rangle$ during the interaction windows.}
\end{figure*}

Now we describe the details of our method. A magnetic field $B_{z}$
parallel to the NV symmetry axis splits the spin triplet of the orbital
ground electronic state of the NV centre. We use two of the three
levels $m_{s}=0,\pm1$ to define an NV electron spin qubit~\cite{doherty2013nitrogen}.
Under strong magnetic fields such that the nuclear Zeeman energies
exceed the perpendicular components, $A_{j}^{\perp}$, of the hyperfine
field $\boldsymbol{A}_{j}$ at the locations of nuclear spins, see
Fig.~\ref{fig:FigSketch}a, the interaction between the NV electron
spin and its surrounding nuclear spins is described by
\[
    H_{\text{int}}= \sigma_{z}\otimes\eta\sum_{j}A_{j}^{\parallel}I_{j}^{z}.
\]
with $\sigma_{z}={|\uparrow_{e}\rangle\langle\uparrow_{e}|}-{|\downarrow_{e}\rangle\langle\downarrow_{e}|}$.
Here we use ${|\uparrow_{e}\rangle}={|m_{s} = +1\rangle}$ as one of the qubit states
while the second qubit state may be $|\downarrow_{e}\rangle=|m_s=0\rangle$, when
$\eta=1/2$, or $|\downarrow_{e}\rangle=|-1\rangle$ with $\eta=1$ (see Supplementary
Information). For each nuclear spin, $A_{j}^{\parallel}$ denotes the component of $\boldsymbol{A}_{j}$ parallel to the nuclear spin
quantisation axes (see Fig.~\ref{fig:FigSketch}a,b). The nuclear precession frequencies
are shifted by $A_{j}^{\parallel}$, which can be used to address
individually nuclear spins of the same species (homonuclear spins).

An initial superposition state of the electron spin $|\psi_{e}\rangle = c_{\uparrow}|\uparrow_{e}\rangle+c_{\downarrow}|\downarrow_{e}\rangle$
loses its coherence because of the electron-nuclear coupling $H_{\text{int}}$.
This effect can be removed by the Hahn echo. In our
case, a microwave $\pi$ pulse exchanges the states $|\uparrow_{e}\rangle \leftrightarrow|\downarrow_{e}\rangle$ and effectively reverses $H_{\text{int}}
\rightarrow -H_{\text{int}}$. When the evolutions before and after the $\pi$
pulse have the same duration $\tau$, entanglement from all the nuclear spins
is erased, preserving the electron spin coherence.

In order to preserve the coupling with a target spin, we apply rf-driving at the
precession frequency of the target spin during a delay window. By flipping
the target spin $j$ by an angle $\theta_{\text{rf}}=\pi$, we selectively rephase
its interaction with the electron spin and realize a quantum gate
\[
|\uparrow_{e}\rangle\langle\uparrow_{e}|\otimes U_{+}+|\downarrow_{e}\rangle\langle\downarrow_{e}|\otimes U_{-}
\]
with $U_{\pm}=\exp\left(\mp2i\eta\tau A_{j}^{\parallel}I_{j}^{z}\right)$, up to
single spin rotations. This leads to the possibility of creating entanglement
between the electron spin and selected nuclear spins which forms a key ingredient
of our method. Note that it is not essential that $\theta_{\text{rf}}=\pi$ as
almost any choice of $\theta_{\text{rf}}$ will generate entanglement with the target and
lead to an observable loss of coherence in the electron spin due to entanglement
with the target nucleus. Additionally, more than one nuclear spins can be addressed 
in the same delay window by using rf driving with different frequencies.

We will describe two strategies to achieve selective nuclear spin control
in the delay window. In the first one we use DD techniques to suppress electron-nuclear
interactions for protection of both the NV electron coherence and the rf nuclear spin control (Fig.~\ref{fig:FigSketch}e).
Consider an NV centre in a diamond sample with natural $^{13}\text{C}$ abundance
(1.1\%) and initialise the NV-center in the equally weighted superposition state
$|\psi_e\rangle$. Then the population signal $P = (1+L)/2$ is directly related
to the observable of NV electron coherence~\cite{zhao2012sensing}, $L = |\langle\downarrow_{e}|\psi_e\rangle\langle
\psi_e|\uparrow_{e}\rangle|$. Fig.~\ref{fig:FigSpectrum}a,b
shows the coherence when scanning the frequency of an rf-pulse with length
$t_{\text{rf}}\approx1$ ms in the delay window protected by 100-pulse Carr-Purcell
(CP) sequences~\cite{mehring2012principles}. When the rf frequency matches one of the
nuclear precession frequencies $\omega_{j}=\omega_{^{13}\text{C}} - A_{j}^{\parallel}/2$
under strong magnetic fields (with $\omega_{^{13}\text{C}}$ denoting the bare Larmor
frequency of $^{13}\text{C}$), we observe a coherence $L = L(A_{j}^{\parallel},
\theta_{\text{rf}})$ where the single spin contribution $L(A_{j}^{\parallel},
\theta_{\text{rf}})=[(1-\cos\theta_{\text{rf}})\cos(2\eta A_{j}^{\parallel}\tau)
+1+\cos\theta_{\text{rf}}]/2$ (see Supplementary Information). When there are $p$
spins with the same $\omega_{j}$, the coherence signal becomes $L=[L(A_{j}^{\parallel},
\theta_{\text{rf}})]^{p}$. The one-to-one correspondence
between $L$ and $\omega_{j}$ (see the dashed lines on Fig.~\ref{fig:FigSpectrum}
where $p=1$ for C1 and C3 and $p=2$ for C2) and the coherence patterns (see
Fig.~\ref{fig:FigSpectrum}c,d) easily identifies the number of nuclear spins
in a dip, even when they are not resolved in the spectrum. Note that the signal
is strong even when $A_{j}^{\perp}=0$, which contrasts previous
methods~\cite{kolkowitz2012sensing,taminiau2012detection,zhao2012sensing,casanova2015robust,wang2015positioning,ma2016scheme}
requiring relatively large $A_{j}^{\perp}$.

We can enhance the interaction and signal by transferring the electronic
state $|\downarrow_{e}\rangle=|0\rangle$ to $|\downarrow_{e}\rangle=|-1\rangle$
during the interaction windows {[}$(t_{1},t_{2})$ and $(t_{5},t_{6})$
in Fig.~\ref{fig:FigSketch}{]}, as shown in Fig.~\ref{fig:FigSpectrum}b.
Fig.~\ref{fig:FigSpectrum}b also shows that changing the rotation
angle $\theta_{\text{rf}}$ does not affect the locations of signal
dips, demonstrating an intrinsic robustness of our method in nuclear
spin detection.

Alternatively, during the delay window we may store the electron spin state in a
long-lived nuclear spin (see Fig.~\ref{fig:FigSketch}f). In Fig.~\ref{fig:FigAncilla}a,
we show the NV electron spin population signal when it addresses an isolated
$^{13}\text{C}$ spin, by using an initially polarised memory qubit  (see Methods). Because the electron spin is polarised to $m_{s}=+1$ during the delay window after a
SWAP gate (see Supplementary Information), it is not necessary to protect electron coherence.
Additionally the shifts of the nuclear spin precession frequencies (now
$\omega_{j}=\omega_{^{13}\text{C}}-A_{j}^{\parallel}$ under strong magnetic fields)
are larger than those achieved by the method described earlier or by DD~\cite{kolkowitz2012sensing,taminiau2012detection,zhao2012sensing,casanova2015robust,wang2015positioning}. The electron
spin relaxation creates magnetic noise on the nuclear spins and can reduce the signal
contrast~\cite{maurer2012room}. However, at low temperatures, the relaxation time
$T_{1}$ of NV electron spin can reach minutes~\cite{yang2015high}. When $T_{1}\gg
t_{\text{rf}}$ the effects of electron spin relaxation can be neglected. Additionally, 
we can protect the memory qubit against the electron-relaxation noise for example
by strong driving~\cite{maurer2012room}. Even for unpolarised memory qubit, the 
signal contrast is still large, showing that our method is insensitive to initialisation 
error of the qubit memory (Fig.~\ref{fig:FigAncilla}a).

Structural information of coupled spin clusters can also be observed
by our method. The dynamics of two coupled spins under a strong magnetic
field is characterised by the dipolar coupling strength $d_{j,k}$
and the difference $\delta_{j,k}=A_{j}^{\parallel}-A_{k}^{\parallel}$
between the hyperfine components~\cite{zhao2011atomic}. A pair of $^{13}\text{C}$
spins with $d_{j,k}\approx\delta_{j,k}$ close to the NV centre can
modulate the NV coherence~\cite{zhao2011atomic,shi2014sensing},
in the absence of rf control as shown in Fig.~\ref{fig:FigAncilla}b.
Spin pairs with $|\delta_{j,k}|\gg|d_{j,k}|$ or $|\delta_{j,k}|\ll|d_{j,k}|$,
which could be useful quantum resources, were regarded as unobservable,
because they form stable pseudo-spins which have negligible effects
on electron coherence~\cite{zhao2011atomic}. But using our method,
we can detect, identify, and control those spin pairs close to the
NV centre (see Supplementary Information for details).

To address and control nuclear spins individually in a coupled cluster
the internuclear dipolar coupling needs to be suppressed. Internuclear
interactions also reduce the Hahn-echo electron coherence times and, hence,
the achievable interaction times $\tau$ (available values $\tau\sim0.5$ ms
for natural abundance of $^{13}\text{C}$ and can be increased using lower
abundance)~\cite{cramer2015repeated,zhao2012decoherence}. To solve the problem
of spin interactions, we use the Lee-Goldburg (LG) off-resonance decoupling~\cite{mehring2012principles} (see Methods).
When the LG decoupling field is tuned such that $\sqrt{2}\Delta_{\text{LG}}\gg d_{jk}$,
the dipolar coupling between nuclear spins are suppressed~\cite{wang2015positioning,cai2013large},
giving rise to the effective interaction Hamiltonian $H_{\text{int}}\approx\eta
\sum_{j}A_{j}^{\parallel}\cos\gamma_{j}\sigma_{z}\tilde{I}_{j}^{z}/\sqrt{3}$
with $\tilde{I}_{j}^{z}$ the nuclear spin operators projected along
effective rotating axes (see Supplementary Information). Fig.~\ref{fig:FigAncilla}b
demonstrates the effect of LG decoupling with $\Delta_{\text{LG}}=2\pi\times20$
kHz. The suppression of the internuclear interactions allows us to achieve
much longer interaction times $\tau$ and rf pulse lengths $t_{\text{rf}}$
for single spin addressing. The improved spectral resolution $\sim1/t_{\text{rf}}$
leads to an increase of the number of individually addressable spins (see Fig.~\ref{fig:FigAncilla}e).

In the case that there are other significant decoherence sources that are
acting on the NV electron spin, our method can be combined with DD to further
protect the NV coherence. Applying CP sequences with an inter $\pi$-pulse interval
$\pi/\omega_{\text{DD}}$ during the interaction windows, noise with frequencies
slower than $\omega_{\text{DD}}$ is suppressed, and at the same time the electron
spin couples to nuclear spins through the interactions $2\eta/(k\pi)A_{j}^{\perp}
\sigma_{z}I_{j}^{\varphi_{j}}$ when the nuclear precession frequencies $\omega_{j}$
are resonant with $k\omega_{\text{DD}}$ ($k$ being odd integers)~\cite{taminiau2012detection,casanova2015robust,wang2015positioning}.
The nuclear spin operators $I_{j}^{\varphi_{j}}=I_{j}^{x}\cos\varphi_{j}+I_{j}^{y}\sin\varphi_{j}$
depends on the azimuthal angles $\varphi_{j}$ of nuclear spins relative
to the magnetic field direction.
Nuclear spins unresolvable by the CP sequences may nevertheless have different 
precession rates. To ensure the same effective Hamiltonian during the interaction 
windows, we apply a two-pulse CP sequence on the nuclear spins during the delay 
window to remove this inhomogeneity. Then adding a weak rf drive during the delay 
window allows us to address the target spins with high spectral resolutions. 
Additionally, the scheme allows us to measure the spin directions $\varphi_{j}$, 
because the rf driving has negligible effects when the azimuthal angle (phase) of 
rf control $\phi_{\text{rf}}=\varphi_{j}$ or $\varphi_{j}+\pi$ (see Fig.~\ref{fig:FigAncilla}c 
and Supplemental Information). Note that we can combine LG decoupling with DD using recently proposed protocols
\cite{wang2015positioning,casanova2016noise}.

Our method allows to improve the spectral resolution beyond the limit
set by the room-temperature electron $T_{1}$, using optical illumination
that has been demonstrated to prolong the room-temperature coherence
time of nuclear spin memory over one second ($\sim267$ times of $T_{1}$)~\cite{maurer2012room}.
To demonstrate the idea, we simulate the application of optical illumination
and rf driving during the delay window to detect a proton spin placed
4 nm away from the NV centre with interaction times $\tau=100$ $\mu$s.
A delay time $t_{\text{rf}}\approx80$ ms used in Fig.~\ref{fig:FigAncilla}d
already provide enough frequency resolution to detect chemical shift
of $\sim1$ ppm for the applied magnetic field $B_{z}\approx0.467$
T. We can apply LG decoupling when there are more target spins and
internuclear interactions. In addition, we can use DD to protect the
interaction windows from noise for extending the interaction time
$\tau$. Electron spin coherence time of shallow NV centre has reported
values of $\sim1$ ms using continuous DD (spin lock)~\cite{lovchinsky2016nuclear}.

In summary, we have proposed a method to address and control nuclear
spins which were regarded as unresolvable. The method significantly
increases the ability of detection and coherence control of nuclear
spins and has applications in quantum information processing as well
as analysis of chemical shifts and dynamics of spin clusters. The
method is general and can be applied to other electron-nuclear spin
systems~\cite{widmann2015coherent,dehollain2016bell}.

\noindent \textbf{Methods}

\noindent The memory nuclear
spin qubit can be initialized by swapping the initialized NV electron
qubit state to the memory spin or by using dynamical nuclear polarization.
Details on the swap operations are presented in Supplemental Information.

The LG decoupling field~\cite{lee1963nuclear,cai2013large,wang2015positioning} can remain turned on for the entire duration of our protocol, including NV electron spin initialization and readout, because the frequency of rf decoupling field is far off-resonance to the transition frequencies of the NV electron spin. This allows for the rf decoupling field to be applied by external coils and resonators to avoid possible heating on the diamond sample. The LG decoupling with $\Delta_{\text{LG}}=2\pi \times 20$ kHz requires a rf field's amplitude to be
much smaller than the values of $\sim0.1$ T in the control fields reported in refs.~\onlinecite{michal2008two,fuchs2009gigahertz}.

\noindent \textbf{Acknowledgements}\\
\noindent This work was supported by the Alexander von Humboldt Foundation,
the ERC Synergy grant BioQ, the EU projects DIADEMS, SIQS and EQUAM
as well as the DFG via the SFB TRR/21. We thank Thomas
Unden and Fedor Jelezko for discussions. Simulations were performed on 
the computational resource bwUniCluster funded by the Ministry of Science, 
Research and the Arts Baden-W\"urttemberg and the Universities of the State 
of Baden-W\"urttemberg, Germany, within the framework program bwHPC.
\\ \\
\noindent \textbf{Author contributions}\\
Z.Y.W., J.C., and M.B.P. conceived the idea. Z.Y.W. carried out the simulations and analytical work  with input from J.C. and M.B.P. All authors discussed extensively on the results and contributed to the manuscript.  
\\ \\
\noindent \textbf{Additional information}\\
The authors declare no competing financial interests. Correspondence and requests for materials should be addressed to Z.Y.W. (zhenyu3cn@gmail.com), J.C. (jcasanovamar@gmail.com), or M.B.P. (martin.plenio@uni-ulm.de).

\pagebreak
\clearpage
\widetext
\begin{center}
\textbf{\large Supplementary Information for ``Delayed entanglement echo for individual
control of a large number of nuclear spins''}
\end{center}
\setcounter{equation}{0}
\setcounter{figure}{0}
\setcounter{table}{0}
\setcounter{page}{1}
\makeatletter
\renewcommand{\theequation}{S\arabic{equation}}
\renewcommand{\thefigure}{S\arabic{figure}}
\renewcommand{\bibnumfmt}[1]{[S#1]}
\renewcommand{\citenumfont}[1]{S#1}

\tableofcontents{}

\section{Hamiltonian of NV centre and nuclear spins}

Under a magnetic field $\boldsymbol{B}=B_{z}\hat{z}$ along the NV
symmetry axis, the Hamiltonian of NV centre electron spin and its
nuclear environment reads ($\hbar=1$)
\begin{equation}
    H=H_{\text{NV}}+H_{\text{nZ}}+H_{\text{hf}}+H_{\text{nn}}.
\end{equation}
Here $H_{\text{NV}}=DS_{z}^{2}-\gamma_{e}B_{z}S_{z}$ is the electron
spin Hamiltonian with the spin operator $S_{z}=\sum_{m_{s}=\pm1,0}m_{s}|m_{s}\rangle\langle m_{s}|$,
the ground state zero field splitting $D\approx2\pi\times2.87$ GHz,
and $\gamma_{e}=-2\pi\times2.8$ MHz/G the electron spin gyromagnetic
ratio~\cite{Sdoherty2013nitrogen}. The nuclear Zeeman Hamiltonian $H_{\text{nZ}}=-\sum_{j}\gamma_{j}\boldsymbol{B}\cdot\boldsymbol{I}_{j}$,
where $\gamma_{j}$ is the nuclear gyromagnetic ratio and $\boldsymbol{I}_{j}$
is the spin operator for the $j$-th nuclear spin. The dipole-dipole
interactions between nuclear spins are
\begin{equation}
    H_{\text{nn}}=\sum_{j>k}\frac{\mu_{0}}{4\pi}\frac{\gamma_{j}\gamma_{k}}{|\boldsymbol{r}_{j,k}|^{3}}\left[\boldsymbol{I}_{j}\cdot\boldsymbol{I}_{k}-\frac{3(\boldsymbol{I}_{j}\cdot\boldsymbol{r}_{j,k})(\boldsymbol{r}_{j,k}\cdot\boldsymbol{I}_{k})}{|\boldsymbol{r}_{j,k}|^{2}}\right],\label{eq:Hnn}
\end{equation}
with $\mu_{0}$ being the vacuum permeability, $\boldsymbol{r}_{j,k}=\boldsymbol{r}_{j}-\boldsymbol{r}_{k}$
the difference between the $k$-th and $j$-th nuclear positions.
Typically the electron-nuclear flip-flop terms in the hyperfine interaction
$H_{\text{hf}}$ are suppressed by the large energy mismatch between
electron and nuclear spins, giving $H_{\text{hf}}=S_{z}\sum_{j}\boldsymbol{A}_{j}\cdot\boldsymbol{I}_{j}$
under the secular approximation. However, for strong hyperfine interactions
the virtual flips of the electron spin could cause observable effects,
an aspect that we will discuss later (see Sec.~\ref{sub:Storage-to-N}). For
nuclear spins not too close to the NV centre, the hyperfine interaction
takes the dipolar form and the hyperfine field
\begin{equation}
    \boldsymbol{A}_{j}=\frac{\mu_{0}}{4\pi}\frac{\gamma_{e}\gamma_{j}}{|\boldsymbol{r}_{j}|^{3}}\left(\hat{z}-\frac{3\hat{z}\cdot\boldsymbol{r}_{j}\boldsymbol{r}_{j}}{|\boldsymbol{r}_{j}|^{2}}\right).\label{eq:Aj}
\end{equation}

Because the total Hamiltonian under secular approximation commutes
with $H_{\text{NV}}$, we simply remove it by going to the rotating
frame with respect to $H_{\text{NV}}$. We choose two of the three triplet states
as the qubit basis states for the NV electron spin. The Hamiltonian
becomes
\begin{equation}
    H_{\eta}=\eta\sigma_{z}\sum_{j}\boldsymbol{A}_{j}\cdot\boldsymbol{I}_{j}+H_{\text{n},\eta}+H_{\text{nn}}.
\end{equation}
In the manifold of the electron spin levels $|\uparrow_{e}\rangle=|+1\rangle$
and $|\downarrow_{e}\rangle=|0\rangle$, we have the coupling constant
$\eta=1/2$, while for the electron spin levels $|\uparrow_{e}\rangle=|+1\rangle$
and $|\downarrow_{e}\rangle=|-1\rangle$, $\eta=1$. The nuclear Hamiltonian
describing nuclear precession reads
\begin{equation}
    H_{\text{n},\eta}=-\sum_{j}(\gamma_{j}\boldsymbol{B}-c_{\eta}\boldsymbol{A}_{j})\cdot\boldsymbol{I}_{j}\equiv-\omega_{j}\hat{\omega}_{j}\cdot\boldsymbol{I}_{j},
\end{equation}
where the unit vectors $\hat{\omega}_{j}$ denote the directions of
$\gamma_{j}\boldsymbol{B}-c_{\eta}\boldsymbol{A}_{j}$ with $c_{\eta}=1/2$
when $\eta=1/2$ and $c_{\eta}=0$ if $\eta=1$. For the case of $H_{1}$ ($\eta=1$),
the electron-nuclear coupling is stronger and the nuclear precession
frequencies $\omega_{j}=\gamma_{j}B_{z}$ are the bare nuclear Larmor
frequencies. While for the case of $H_{\frac{1}{2}}$ ($\eta=1/2$), the electron-nuclear
coupling is weaker and the precession frequencies $\omega_{j}=|\gamma_{j}\boldsymbol{B}-\frac{1}{2}\boldsymbol{A}_{j}|$
are shifted by the hyperfine field at the positions of the nuclear
spins.

In the rotating frame of nuclear spin precession $H_{\text{n},\eta}$, the interaction Hamiltonian
$\eta\sigma_{z}\sum_{j}\boldsymbol{A}_{j}\cdot\boldsymbol{I}_{j}$
becomes~\cite{Scasanova2015robust}
\begin{equation}
    H_{\text{int}}=\eta\sigma_{z}\sum_{j}\left[\boldsymbol{A}_{j}^{x}\cos(\omega_{j}t)+\boldsymbol{A}_{j}^{y}\sin(\omega_{j}t)+\boldsymbol{A}_{j}^{z}\right]\cdot\boldsymbol{I}_{j},\label{eq:Hint}
\end{equation}
with
\begin{eqnarray}
    \boldsymbol{A}_{j}^{x} & \equiv & \boldsymbol{A}_{j}-\boldsymbol{A}_{j}^{z},\\
    \boldsymbol{A}_{j}^{y} & \equiv & \hat{\omega}_{j}\times\boldsymbol{A}_{j},\\
    \boldsymbol{A}_{j}^{z} & \equiv & \boldsymbol{A}_{j}\cdot\hat{\omega}_{j}\hat{\omega}_{j}.
\end{eqnarray}
The hyperfine components have the strengths $|\boldsymbol{A}_{j}^{x}|=|\boldsymbol{A}_{j}^{y}|=A_{j}^{\perp}$
and $|\boldsymbol{A}_{j}^{z}|=A_{j}^{\parallel}$. The time-dependent
terms in Eq.~(\ref{eq:Hint}) do not commute with the nuclear precession
$H_{\text{n},\eta}$.

Under a strong magnetic field $B_{z}\gg A_{j}^{\perp}$, $\hat{\omega}_{j}\approx\hat{z}$.
The nuclear spin flips are suppressed, giving
\begin{equation}
    H_{\text{int}}\approx\eta\sigma_{z}\sum_{j}\boldsymbol{A}_{j}^{z}\cdot\boldsymbol{I}_{j}=\eta\sigma_{z}\sum_{j}A_{j}^{\parallel}I_{j}^{z}.\label{eq:HintZZ}
\end{equation}
If we apply Lee-Goldburg (LG) off-resonance control~\cite{Slee1965nuclear}, we can achieve similar Hamiltonians~\cite{Swang2015positioning,Scai2013large}
\begin{equation}
    H_{\text{int}}^{\text{LG}}\approx\eta\sigma_{z}\sum_{j}A_{j}^{\parallel}\cos\gamma_{j}\tilde{I}_{j}^{z},\label{eq:HintZZLG}
\end{equation}
where $\tilde{I}_{j}^{z}=\hat{\nu}_{j}\cdot\boldsymbol{I}_{j}$ with
$\hat{\nu}_{j}$ the unit vector denoting the nuclear precession in
the frame of LG control. The projection factor $\cos\gamma_{j}=\hat{\omega}_{j}\cdot\hat{\nu}_{j}\approx1/\sqrt{3}$.

The interaction Hamiltonian Eq.~(\ref{eq:HintZZ}) commutes with the
nuclear precession. Similarly, Eq.~(\ref{eq:HintZZLG}) commutes with
the nuclear precession around $\hat{\nu}_{j}$ in the frame of LG control.
Combined with the delay entanglement control described in the main text, we
can keep only terms on the target spins in $H_{\text{int}}$
or $H_{\text{int}}^{\text{LG}}$. The effective electron-nuclear
interactions by delay entanglement control do not broaden the nuclear
precession frequencies for addressing.

\section{Spin addressing by dynamical decoupling}

\subsection{Effective interaction Hamiltonians under dynamical decoupling\label{sub:DD-Effective-interaction-Hamiltonia}}

Nuclear spins can be addressed by dynamical decoupling (DD)~\cite{Skolkowitz2012sensing,Staminiau2012detection,Szhao2012sensing,Slondon2013detecting,Smkhitaryan2015highly,Scasanova2015robust,Swang2015positioning}.
The DD pulses flip the NV electron qubit. After application of $n$
$\pi$ pulses, $\sigma_{z}\rightarrow F(t)\sigma_{z}$ with the modulation
function $F(t)=(-1)^{n}$. We consider periodic sequences with $F(t)=F(t+2\pi/\omega_{\text{DD}})$
in this work. The interaction Hamiltonian Eq.~(\ref{eq:Hint}) becomes
\begin{equation}
H_{\text{int}}=\eta F(t)\sigma_{z} \sum_{j}\left[\boldsymbol{A}_{j}^{x}\cos(\omega_{j}t)+\boldsymbol{A}_{j}^{y}\sin(\omega_{j}t)+\boldsymbol{A}_{j}^{z}\right]\cdot\boldsymbol{I}_{j}.
\end{equation}
This instantaneous-pulse control changes the electron-nuclear dynamics~\cite{Staminiau2012detection,Staminiau2014universal,Sliu2013noise}.
To get insight on nuclear spin sensing by DD pulse sequences, we expand
the modulation function in a Fourier series,
\[
F(t)=\sum_{k\geq1}[f_{k}^{\text{s}}\cos(k\omega_{\text{DD}}t)+f_{k}^{\text{a}}\sin(k\omega_{\text{DD}}t)],
\]
using that DD pulses have been designed to remove static noise and
hence there is no static term in the Fourier series. For periodic
symmetric sequences $f_{k}^{\text{a}}=0$. The frequency $\omega_{\text{DD}}$
characterises the flipping rate of the NV electron qubit. For example,
for the traditional Carr-Purcell (CP) sequence~\cite{Scarr1954effects}
and its variations~\cite{Smeiboom1958modified,Smaudsley1986modified,Sgullion1990new}
having a time interval $\tau_{\text{CP}}$ between successive $\pi$
pulses, $\omega_{\text{DD}}=\pi/\tau_{\text{CP}}$, $f_{k}^{\text{s}}=4(k\pi)^{-1}\sin(k\pi/2)$,
and $f_{k}^{\text{a}}=0$. The expansion coefficients can be tuned
by adaptive XY (AXY) sequences~\cite{Scasanova2015robust}.

A nuclear spin with the precession frequency $\omega_{n}$ can be
addressed by resonance to the $k_{\text{DD}}$ harmonic of the driving
rate, that is, $\omega_{n}=k_{\text{DD}}\omega_{\text{DD}}$. With
the additional conditions (with $j\neq n$) $|\gamma_{j}B_{z}|\gg k_{\text{DD}}|\boldsymbol{A}_{j}|$
and
\begin{eqnarray}
|\omega_{n}-\omega_{j}| & \gg & |f_{k_{\text{DD}}}A_{j}^{\perp}|,\label{eq:RWAWeekFdd}
\end{eqnarray}
we have single spin addressing under periodic symmetric sequences
$H_{\text{int}}\approx(\eta/2) f_{k_{\text{DD}}}^{\text{s}}A_{j}^{\perp}\sigma_{z}I_{j}^{x}$~\cite{Scasanova2015robust}.
Similar addressing Hamiltonians ($H_{\text{int}}^{\text{LG}}\propto \sigma_{z}\tilde{I}_{j}^{x}$ with $\tilde{I}_{j}^{x}$ a spin operator projected perpendicular to $\hat{\nu}_{j}$) can be achieved under LG control~\cite{Swang2015positioning,Scasanova2016noise}.

Nuclear spins can also be addressed by continuous DD~\cite{Slondon2013detecting}.
In the rotating frame of a constant microwave driving $\Omega_{e}\sigma_{x}/2$
with the Rabi frequency $\Omega_{e}$ (the frequency of nuclear spin
precession in the spin-lock frame), the Pauli operator of NV electron
qubit transforms as $\sigma_{z}\rightarrow\sigma_{z}\cos(\Omega_{e}t)+\sigma_{y}\sin(\Omega_{e}t)$.
The driving rate of the electron spin is $\omega_{\text{DD}}=\Omega_{e}$.
When $\Omega_{e}$ is on-resonance to the nuclear spin precession
frequency $\omega_{j}$, that is, $\Omega_{e}=\omega_{j}$, we have
the addressing Hamiltonian $H_{\text{int}}\approx(\eta/2) A_{j}^{\perp}(\sigma_{z}I_{j}^{x}+\sigma_{y}I_{j}^{y})$
when $|\gamma_{j}B_{z}|\gg|\boldsymbol{A}_{j}|$ and $|\omega_{n}-\omega_{j}|\gg A_{j}^{\perp}$
for $j\neq n$.

\subsection{Shortcomings of spin addressing by dynamical decoupling}

\begin{figure*}
\includegraphics[clip,width=0.9\textwidth]{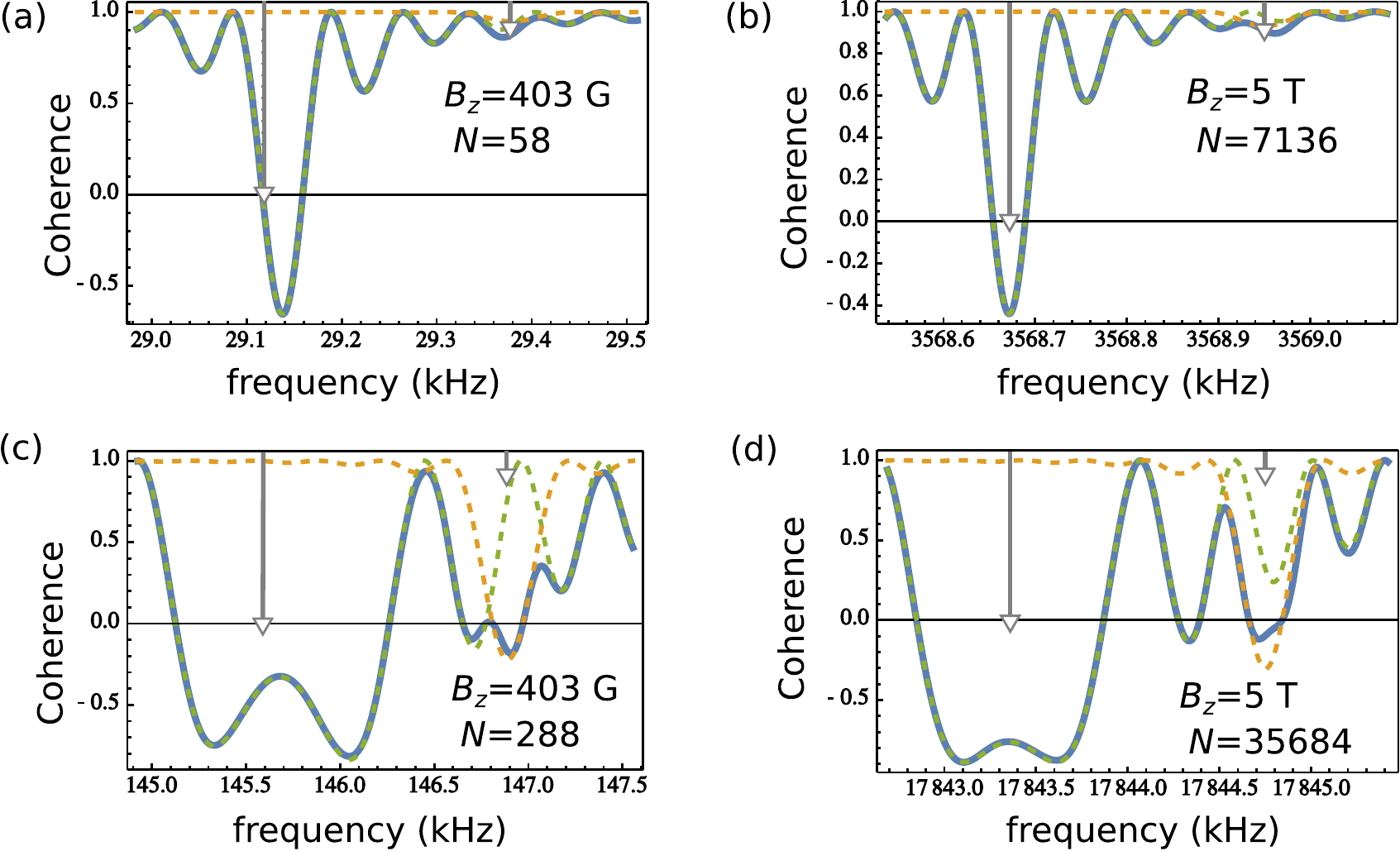}\caption{\label{fig:smFigSpectrumCPMG} \textbf{Coherence signals obtained
by CP sequences.}  Coherence
of NV electron spin (blue solid line) as a function of DD frequency
$\omega_{\text{DD}}=\pi/\tau_{\text{CP}}$ (with $\tau_{\text{CP}}$
the pulse intervals of CP sequences), under the control of $N$-pulse
CP sequences at a magnetic field $B_{z}$. The CP sequences have durations
of $\sim1$ ms using the pulse numbers and magnetic fields indicated
on the figures. The green and yellow dashed lines are the signals
of single spins. The arrows indicate the frequencies $\omega_{j}/k_{\text{DD}}$
with the lengths proportional to $A_{j}^{\perp}$.
The signals of resonances at $k_{\text{DD}}=15$ in (a) and (b) are dominated by the
nuclear spin with a strong $A_{j}^{\perp}$. While in (c) and (d),
increasing the interactions by using a smaller $k_{\text{DD}}=3$
also increases the interference from the spin with a strong $A_{j}^{\perp}$,
prohibiting the spin addressing on the nuclear spin with a weaker
$A_{j}^{\perp}$.  In (b) and (d), the number of pulses increases
under a stronger magnetic field, compared with the cases in (a) and (c).}
\end{figure*}

\begin{figure*}
\includegraphics[clip,width=0.9\textwidth]{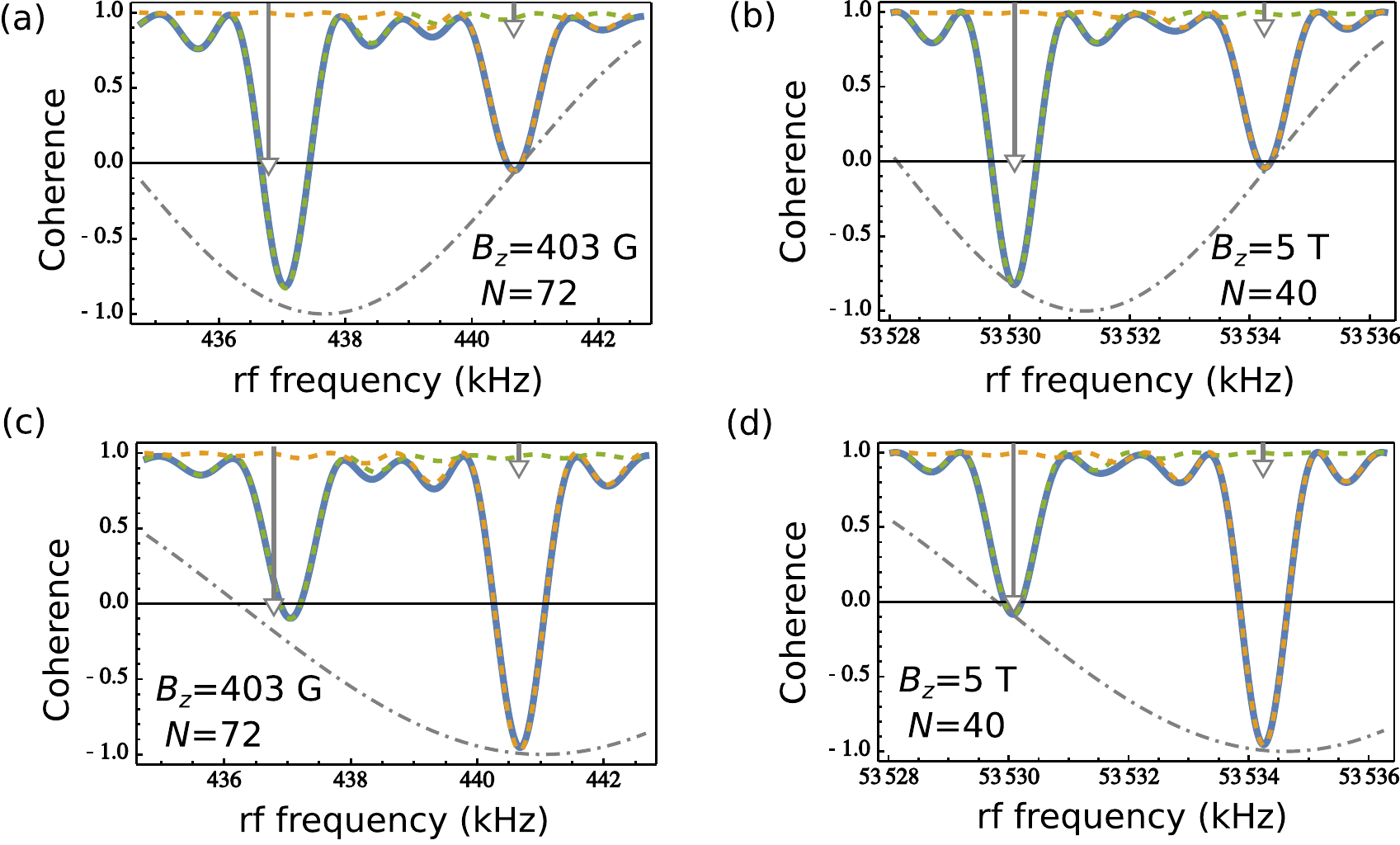}\caption{\label{fig:smFigSpectrumDEE} \textbf{Coherence signals obtained by
delayed entanglement echoes.}
Coherence of NV electron spin (blue solid line) as a function of the
frequency $\omega_{\text{rf}}$ of rf driving at the delay window,
which the duration $t_{\text{rf}}\approx1$ ms and $\theta_{\text{rf}}=\pi$.
The pulse number $N$ of CP sequences used to protect the delay window
are indicated on the figures, as well as the magnetic filed $B_{z}$.
The green and yellow dashed lines are the signals of single spins.
Each of the arrow located at the nuclear precession frequency $\omega_{j}$
has a length proportional to $A_{j}^{\perp}$. The interaction times
$\tau=20~\mu$s for (a) and (b), while a shorter $\tau=13~\mu$s for
(c) and (d). Dash-dotted lines are the curve
$\cos[4\tau(\omega_{\text{rf}}-\omega_{^{13}\text{C}})]$. The qubit levels $m_{s}=\pm1$ are used at the interaction windows. }
\end{figure*}

The addressing by DD has a number of shortcomings that we are going to discuss in the following.

First, the interaction Hamiltonians achieved by DD do not commute with the nuclear precession.
As a consequence, electron-nuclear interactions broaden the nuclear precession frequencies for
addressing (see Figs.~\ref{fig:smFigSpectrumCPMG} (c) and (d)). We need the condition in Eq.~(\ref{eq:RWAWeekFdd})
for individual spin addressing. 
Reducing the effective interaction strengths between NV electron
and nuclear spins improves a lot the spectral resonance by using higher harmonics $k_{\text{DD}}>1$, by 
alternating the phase of Rabi driving~\cite{Smkhitaryan2015highly}, or
by using composite $\pi$ pulses~\cite{Szhao2014dynamical,Scasanova2015robust,Sma2015resolving,Swang2015positioning}.
But the reduced coupling also makes nuclear spins that are not strongly coupled hard to detect and
control (see Figs.~\ref{fig:smFigSpectrumCPMG} (a) and (b)).

Second, because resonances can occur at different harmonics frequencies
$k\omega_{\text{DD}}$, resonance lines from different harmonic branches
can have overlaps and make critical ambiguities in detection and addressing~\cite{Staminiau2012detection,Sloretz2015spurious,Scasanova2015robust}.
The spurious resonances caused by realistic pulse width further complicate
the situation, even making false identification of different nuclear
species (e.g., $^{13}\text{C}$ and $^{1}\text{H}$)~\cite{Sloretz2015spurious}.

Third, the achievable rate $\omega_{\text{DD}}$ of DD sets an upper
limit on the external magnetic field for spin addressing. A strong
magnetic field is the requirement in detecting the chemical shift
of nuclear spins~\cite{Smehring2012principles} and in decoupling
of the nuclear dipole-dipole interactions by rf control~\cite{Slee1965nuclear,Scai2013large,Swang2015positioning}.
In addition, the NV electron coherence can be protected easier under
strong magnetic fields~\cite{Szhao2012decoherence}. However, nuclear
spin precession frequencies $\omega_{j}$ at strong magnetic fields
can be significantly larger than the achievable rate $\omega_{\text{DD}}$
of DD control. For example, the pulse number 35684 required in Fig.~\ref{fig:smFigSpectrumCPMG} (d)
could be too many in experiments. Using resonance branches with large
$k_{\text{DD}}$ can reduce the required control rate $\omega_{\text{DD}}$,
but it also reduces electron-nuclear coupling and narrows spectral
bandwidths (in Fig.~\ref{fig:smFigSpectrumCPMG} (b) the coupling
is too weak to detect the nuclear spin with a weak $A_{j}^{\perp}$
and the bandwidth is about $\sim\omega_{^{13}\text{C}}/k_{\text{DD}}$
for $^{13}\text{C}$ spins).

The delayed entanglement echo technique does not suffer from the above shortcomings
(compare Fig.~\ref{fig:smFigSpectrumDEE} with Fig.~\ref{fig:smFigSpectrumCPMG}),
and provide some additional advantages. First, it does not require
both hyperfine components $A_{j}^{\parallel}$ and $A_{j}^{\perp}$
to be strong. Second, both the electron-nuclear coupling strengths
before and after the delay window are not reduced. In addition, we
can use the levels $m_{s}=\pm1$ to double the interaction strength
(changing $\eta=1/2$ to $\eta=1$), since the nuclear spins are addressed
by the control in the delay window. In contrast, $\eta=1/2$ is necessary
during the whole protocol of spin addressing by standard DD, so that
homonuclear spins feeling different hyperfine fields have different precession frequencies.
Third, our technique allows to simultaneously address more than one nuclear spin by
applying rf driving fields at the frequencies of those spins during a delay window.

\section{Storage of electron states to a memory qubit }

\noindent Here we present more details on storing the electron qubit
states to a nuclear spin memory. During the swap operations, the NV
electron qubit is working in the $m_{s}=0$, $+1$ manifold. Storage
of electron spin state can be realised by SWAP gates. A SWAP gate
\begin{equation}
\text{SWAP}=\sum_{m_{s},m_{n}=0,1}|m_{s}m_{n}\rangle\langle m_{n}m_{s}|
\end{equation}
swap the electron qubit states $m_{s}$ and memory qubit states $m_{n}$.
In the case that relaxations of the electron and nuclear memory qubit
can be neglected during the delay window, we can also use iSWAP gate
\begin{equation}
\text{iSWAP}=\sum_{m_{s},m_{n}=0,1}e^{i(m_{s}+m_{n})^{2}\pi/2}|m_{s}m_{n}\rangle\langle m_{n}m_{s}|,
\end{equation}
which introduces a phase factor $i$ when $m_{s}\neq m_{n}$. Without
relaxations of the electron and nuclear memory qubit, the whole system
including the environment and the memory qubit has the evolution during
the delay window
\begin{equation}
U_{\text{delay}}=\sum_{m_{s},m_{n}=0,1}|m_{s}m_{n}\rangle\langle m_{s}m_{n}|\otimes U_{m_{s},m_{n}},
\end{equation}
where $U_{m_{s},m_{n}}$ are unitary evolution operators of the environment
part. The effect of the iSWAP gates,
\begin{equation}
\text{iSWAP}^{\dagger}U_{\text{delay}}\text{iSWAP}=\sum_{m_{s},m_{n}=0,1}|m_{n}m_{s}\rangle\langle m_{n}m_{s}|\otimes U_{m_{s},m_{n}},
\end{equation}
is the same as using SWAP gates.

We use protected swap gates to suppress decoherence of the NV electron
spin and unwanted electron-nuclear interactions during gate implementation.
Using nuclear spin addressing by DD~\cite{Staminiau2014universal,Scasanova2015robust,Svan2012decoherence,Swang2015positioning}
or the delayed entanglement echo presented in the main text, we implement
the elementary decoherence-protected two qubit gates $u_{z\alpha}=\exp\left(i\frac{\pi}{4}\sigma_{z}I_{\alpha}\right)$
with $\alpha=x,y,z$ as well as single qubit gates for nuclear spins.
Combining the gate \foreignlanguage{english}{$u_{z\alpha}$} with
electron spin rotations, we achieve the gate $u_{\alpha\alpha}=\exp\left(i\frac{\pi}{4}\sigma_{\alpha}I_{\alpha}\right)$
{[}e.g., $u_{yy}=\exp\left(i\frac{\pi}{4}\sigma_{x}\right)u_{zy}\exp\left(-i\frac{\pi}{4}\sigma_{x}\right)${]}.
A swap gate is constructed by $u_{zz}u_{yy}u_{xx}$, where the three
gates $u_{\alpha\alpha}$ commute, while $u_{yy}u_{xx}$ gives rise
to the iSWAP gate.

\subsection{Storage to the intrinsic nitrogen spin\label{sub:Storage-to-N}}

Here we describe the details of implementation of SWAP gates between
the electron qubit and the intrinsic nitrogen spin qubit. For simplicity,
we consider $^{14}\text{N}$, which has $99.636\%$ natural abundance
and a spin $I=1$. The Hamiltonian for the NV electron and the intrinsic
nitrogen spins is
\begin{equation}
H_{\text{NV}}=DS_{z}^{2}-\gamma_{e}B_{z}S_{z}+PI_{z}^{2}-\gamma_{N}B_{z}I_{z}+A^{\parallel}S_{z}I_{z}+A^{\perp}(S_{x}I_{x}+S_{y}I_{y}), \label{eq:ElectronIntNitrogen}
\end{equation}
where $\gamma_{N}=2\pi\times0.308$ $\text{kHz}/\text{G}$. We adopt
the parameters for $^{14}\text{N}$ in NV centres $A^{\perp}=-2\pi\times2.62$
MHz, $A^{\parallel}=-2\pi\times2.162$ MHz, and $P=-2\pi\times4.945$
MHz~\cite{Schen2015measurement}. The flip-flop between electron and
nuclear spins are suppressed by the large energy mismatch. We have
\begin{equation}
H_{\text{NV}}\approx DS_{z}^{2}-\gamma_{e}B_{z}S_{z}+PI_{z}^{2}-\gamma_{N}B_{z}I_{z}+A^{\parallel}S_{z}I_{z}+\sum_{m_{s}=0,\pm1}|m_{s}\rangle\langle m_{s}|h_{m_{s}},
\end{equation}
where the nitrogen operators
\begin{equation}
    h_{+1}=\frac{(A^{\perp})^{2}}{D-\gamma_{e}B_{z}}
    (|0_{N}\rangle\langle0_{N}|+|-1_{N}\rangle\langle-1_{N}|),
\end{equation}
\begin{equation}
    h_{0}=\frac{(A^{\perp})^{2}}{-D+\gamma_{e}B_{z}}|+1_{N}\rangle
    \langle+1_{N}|+\left[\frac{(A^{\perp})^{2}}{-D+\gamma_{e}B_{z}}
    -\frac{(A^{\perp})^{2}}{D+\gamma_{e}B_{z}}\right]|0_{N}\rangle\langle0_{N}|
    -\frac{(A^{\perp})^{2}}{D+\gamma_{e}B_{z}}|-1_{N}\rangle\langle-1_{N}|,
\end{equation}
\begin{equation}
    h_{-1}=\frac{(A^{\perp})^{2}}{D+\gamma_{e}B_{z}}
    (|0_{N}\rangle\langle0_{N}|+|+1_{N}\rangle\langle+1_{N}|),
\end{equation}
describe the energy shifts caused by virtual spin flip-flop processes.

We use the electron-nitrogen coupling to implement the entangled gate $u_{zz}=\exp(i\frac{\pi}{4}\sigma_{z}\sigma_{N,z})$ in a short duration of $0.23~\mu\text{s}$, where $\sigma_{N,z}$ is the Pauli operator of the nitrogen qubit. The iSWAP gate $\text{iSWAP}=u_{yy}u_{xx}$, where $u_{xx}=P_{y}u_{zz}P_{y}^{\dagger}$ 
and $u_{yy}=P_{x}u_{zz}P_{x}^{\dagger}$. Here $P_{\alpha}$ denotes decoherence-protected 
single-qubit $\pi/2$ gates~\cite{Svan2012decoherence} on both nuclear and electron 
spins at the $\alpha$ direction. The SWAP gate can be realized by $\text{SWAP}=e^{-i\pi/4}u_{zz}\text{iSWAP}$.

We simulate the SWAP gate using the protocol and achieve a gate fidelity 
of $F=0.987$ (defined by $F=|\text{Tr}(GU_{g})|/\text{Tr}(GG^{\dagger})$ with $G$ the 
evolution of ideal SWAP gate and $U_{g}$ the actual implementation) by taking into account
the energy shifts on the electron and nitrogen qubits. 
The microwave $\pi$ pulses for the SWAP gate 
have pulse duration of $12.5$ ns, and the field strength for rf $\pi/2$ pulses is $15.53$ G. 
The magnetic field $B_{z}=0.467$ T and the LG decoupling are the same as those used in Fig.~3b in the main text.
In the simulation, we adopt the Hamiltonian Eq.~(\ref{eq:ElectronIntNitrogen}) and 
the electron and nuclear spins feel all the control fields irrespective to their frequencies. 
Because of the high gate fidelity, we use the nitrogen spin levels $m_{N}=0,+1_{N}$ to store the NV 
electron qubit by ideal swap gates in producing the figures in the main text that use the nitrogen spin  as quantum memory.

After the swap operation, the NV electron spin is polarized to the state $|+1\rangle$ 
for the delay window. During this storage, we take into account the energy shifts on 
the nitrogen memory. The energy shifts can also be removed by applying DD (e.g., a Hahn 
echo) on the nitrogen spin.

\subsection{Storage to carbon-13 memory qubits\label{sub:Storage-to-carbon-13}}

\begin{figure*}
\includegraphics[clip,width=0.9\textwidth]{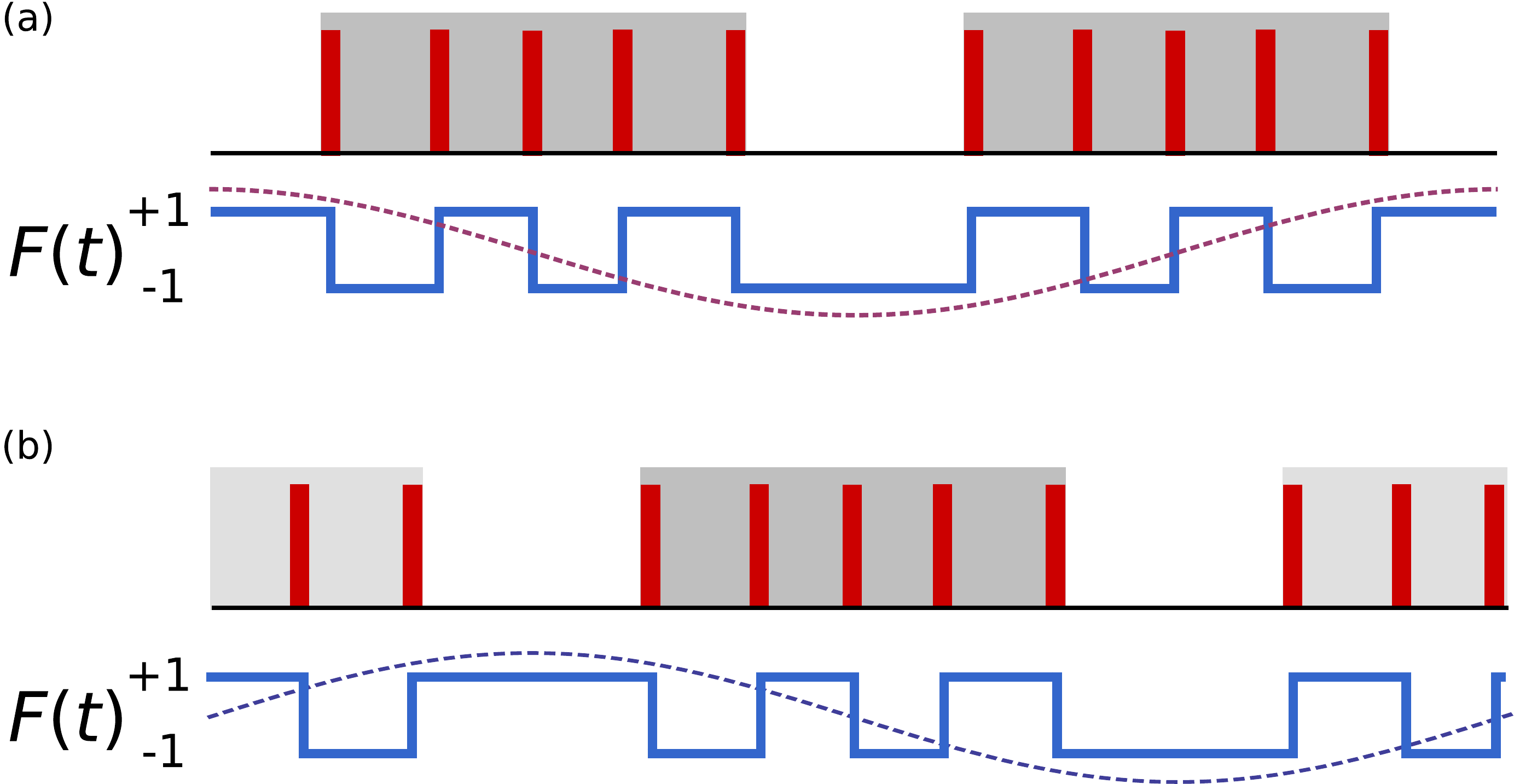}\caption{\label{fig:smFigAXY} 
\textbf{Illustration of AXY pulse sequences
in one period.} (a) Symmetric version and its corresponding modulation function shown 
below. (b) The anti-symmetric counterpart. Each of the shaded areas with a heavier colour 
highlights a composite $\pi$ pulse in the AXY sequences. The dashed lines indicate the 
sinusoidal signals on-resonant to the modulation function $F(t)$.} 
\end{figure*}

We can use AXY sequences~\cite{Scasanova2015robust} to implement $u_{\alpha\alpha}$
gates and swap the NV electron states to a $^{13}\text{C}$ memory.
Compared to traditional sequences, AXY exhibits especially good spin addressability,
 strong robustness against detuning and amplitude errors, and the ability
to continuously tune the effective interactions between NV electron
and nuclear spins~\cite{Scasanova2015robust}. Using a symmetric version
of AXY sequence (see Fig.~\ref{fig:smFigAXY} (a)), we have
the interaction Hamiltonian $H_{\text{int}}^{x}\approx\frac{1}{4}f_{k_{\text{DD}}}^{\text{s}}A_{j}^{\perp}\sigma_{z}I_{j}^{x}$~\cite{Scasanova2015robust}.
Similarly, for anti-symmetric sequences (see Fig.~\ref{fig:smFigAXY} (b)),
we have $H_{\text{int}}^{y}\approx\frac{1}{4}f_{k_{\text{DD}}}^{\text{a}}A_{j}^{\perp}\sigma_{z}I_{j}^{y}$.
We tune $f_{k_{\text{DD}}}^{\text{s}}=f_{k_{\text{DD}}}^{\text{a}}=f_{k_{\text{DD}}}$
and use a time $t_{g}=2\pi/(f_{k_{\text{DD}}}A_{j}^{\perp})$ to implement
the operation
\begin{equation}
\text{iSWAP=}X_{\pi/2}\exp(-iH_{\text{int}}^{y}t_{g})X_{\pi/2}^{\dagger}Y_{\pi/2}^{\dagger}\exp(-iH_{\text{int}}^{x}t_{g})Y_{\pi/2},
\end{equation}
where $X_{\pi/2}$ and $Y_{\pi/2}$ are NV electron $\pi/2$ gates
around the directions $x$ and $y$, respectively. The inverse gate
$\text{iSWAP}^{\dagger}$ can be implemented by the interchanges $X_{\pi/2}\leftrightarrow X_{\pi/2}^{\dagger}$
and $Y_{\pi/2}\leftrightarrow Y_{\pi/2}^{\dagger}$.

Another way to implement the swap gate is by using continuous DD (i.e.,
using spin-locking field). For the addressed nuclear spin with a distinct
precession frequency, we have the effective interaction Hamiltonian
$H_{\text{int}}\approx\frac{1}{2}\eta A_{j}^{\perp}(\sigma_{z}I_{j}^{x}+\sigma_{y}I_{j}^{y})$
under continuous Rabi driving (see Sec.~\ref{sub:DD-Effective-interaction-Hamiltonia}).
An iSWAP gate corresponds to the sequence $\exp\left(-i\frac{\pi}{4}\sigma_{y}\right)\exp\left(-iH_{\text{int}}t_{g}\right)\exp\left(i\frac{\pi}{4}\sigma_{y}\right)$.

In producing the figures in the main text with a $^{13}\text{C}$ memory,
we explicitly implement the swap gate operations by ideal microwave control.

\section{Signals of delayed entanglement echo}

\subsection{Single spins}

The evolution for a target nuclear spin is $U=e^{-i\eta A_{j}^{\parallel}\tau\sigma_{z}I_{j}^{z}}$
(without considering other spins) after an interaction window with
time $\tau$. Then we apply a rotation (say, along the $x$ direction)
with an angle $\theta_{\text{rf}}$ on the target nuclear spin at
the delay window. Following by another interaction window with time
$\tau$ between two $\pi$ pulses on the electron spin (the last $\pi$
pulse is optional, but we keep it for simplicity), we have a total
evolution $U=e^{i\eta A_{j}^{\parallel}\tau\sigma_{z}I_{j}^{z}}e^{-iI_{j}^{x}\theta_{\text{rf}}}e^{-i\eta A_{j}^{\parallel}\tau\sigma_{z}I_{j}^{z}}$
for interaction between the NV electron spin and a target nuclear
spins. Because $k_{B}/\hbar\approx(2\pi)\times21$ GHz/K, the thermal
energies are much larger than the nuclear spin Zeeman energies at
relevant temperatures, and we approximate the thermal state of nuclear
spins by the identity operator up to a normalization factor. Given
the initial electron spin state $|\psi_{e}\rangle=(|\uparrow_{e}\rangle+|\downarrow_{e}\rangle)/\sqrt{2}$,
the population left in the original equal superposition state is
\begin{equation}
P_{\text{NV}}=\frac{1}{2}+\frac{1}{4\mathcal{N}}\text{Tr}(U_{+}U_{-}^{\dagger}+U_{-}U_{+}^{\dagger}),
\end{equation}
with $U_{\pm}=e^{\pm i\eta A_{j}^{\parallel}\tau I_{j}^{z}}e^{-i\theta_{\text{rf}}I_{j}^{x}}e^{\mp i\eta A_{j}^{\parallel}\tau I_{j}^{z}}$
and $\mathcal{N}=2I+1$ the dimension of nuclear spin ($I=1/2$ for
$^{13}\text{C}$).  For spin-$\frac{1}{2}$, we obtain $P_{\text{NV}}=(1+L)/2$
and the coherence $L=L(A_{j}^{\parallel},\theta_{\text{rf}})$, where
the single spin contribution
\begin{equation}
L(A_{j}^{\parallel},\theta_{\text{rf}})=\frac{1}{2}[(1-\cos\theta_{\text{rf}})\cos(2\eta A_{j}^{\parallel}\tau)+1+\cos\theta_{\text{rf}}].\label{eq:L1spin}
\end{equation}
The range of single spin contribution $\cos\theta_{\text{rf}}\leq L(A_{j}^{\parallel},\theta_{\text{rf}})\leq1$.
When there are a number $p$ of nuclear spins at indistinguishable
Larmor frequencies, the coherence $L=L^{p}(A_{j}^{\parallel},\theta_{\text{rf}})$.
For the case of a $\theta_{\text{rf}}=\pi$ rotation, $L(A_{j}^{\parallel},\theta_{\text{rf}})=\cos(2\eta A_{j}^{\parallel}\tau)$.

\begin{figure*}
\includegraphics[clip,width=0.9\textwidth]{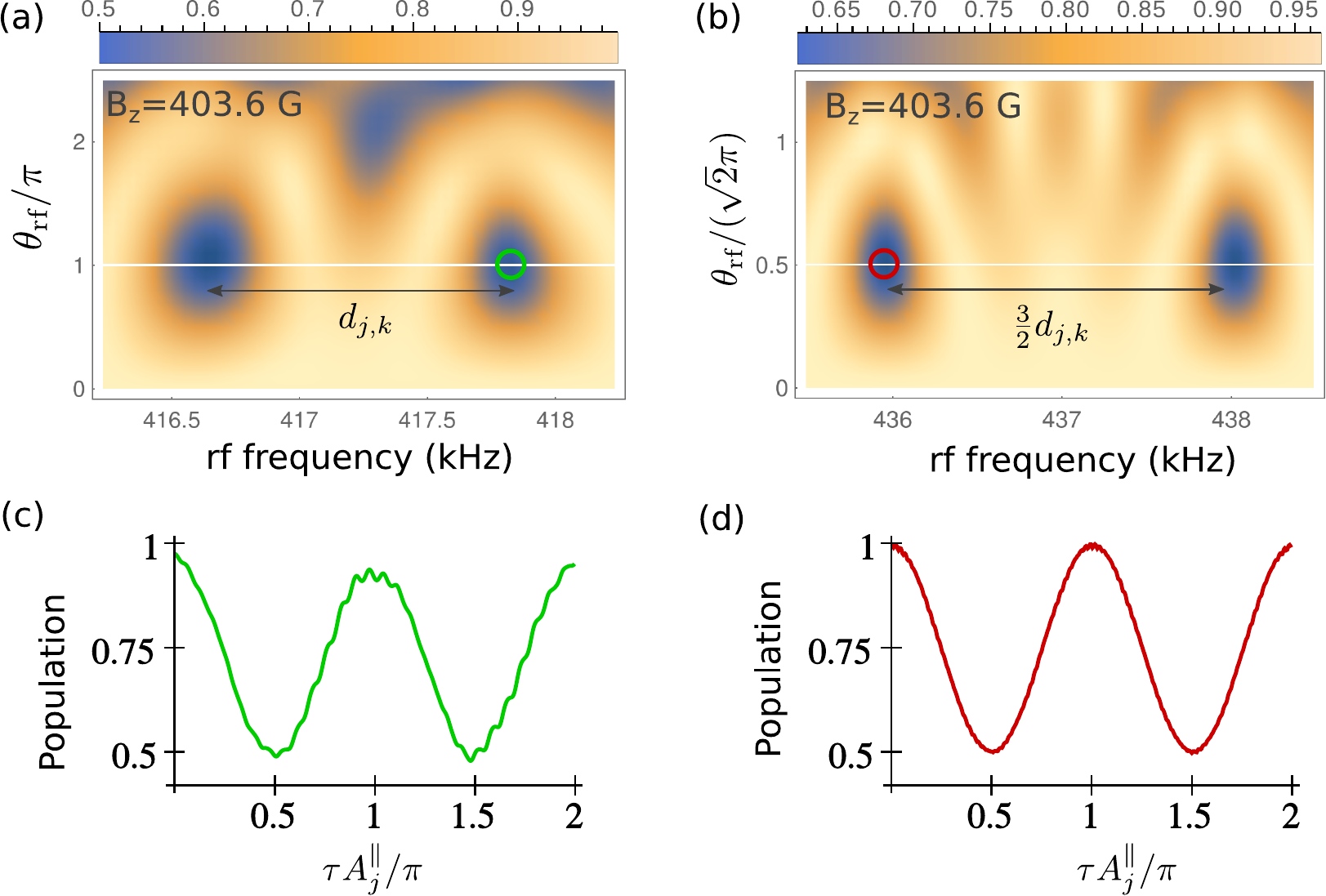}\caption{\label{fig:smFigPair} \textbf{Population signal of coupled spin pairs.}
(a) Signal patterns of two coupled $^{13}\text{C}$ spins
in a C-C bond $\approx1.0$ nm away from the NV centre, using the
interaction time $\tau=10$ $\mu$s and rf pulse length $t_{\text{rf}}\approx2$ ms.
The dipolar coupling strength $d_{j,k}\approx2\pi\times1.37$ kHz.
The hyperfine components of the two spins $A_{j}^{\parallel}\approx2\pi\times15.6$
kHz and $A_{k}^{\parallel}\approx2\pi\times22.5$ kHz. (b)
Same as (a) but for a spin pair located $\approx1.55$ nm away
from the NV centre and $\tau=50$ $\mu$s, and the hyperfine components
of the two spins $A_{j}^{\parallel}\approx-2\pi\times5.00$ kHz and
$A_{k}^{\parallel}\approx-2\pi\times4.96$ kHz have similar strengths.
The parameters used in (c) and (d) are indicated by
the circles in (a) and (b), respectively.}
\end{figure*}

\subsection{Coupled spin pairs}

Under strong magnetic field, there are three types of spin pairs~\cite{Szhao2011atomic},
according to their relative amplitudes between the dipolar coupling
strength
\begin{equation}
d_{j,k}=\frac{\mu_{0}}{4\pi}\frac{\gamma_{j}\gamma_{k}}{|\boldsymbol{r}_{j,k}|^{3}}\left[1-3(\hat{z}\cdot\boldsymbol{r}_{j,k}/|\boldsymbol{r}_{j,k}|)^{2}\right],
\end{equation}
and the difference of the hyperfine components $\delta_{j,k}=A_{j}^{\parallel}-A_{k}^{\parallel}$.
Spin pairs with $\delta_{j,k}$ and $d_{j,k}$ the same order of strengths,
which we call type-s, can be detected by DD because it modulates the
electron spin coherence~\cite{Szhao2011atomic,Sshi2014sensing}. This
type of spin pairs can be detected by our method without application
of rf driving at the delay window (i.e., $\theta_{\text{rf}}=0$). Other types of spin pairs, type-h
($|\delta_{j,k}|\gg|d_{j,k}|$) and type-d ($|d_{j,k}|\gg|\delta_{j,k}|$),
were regarded as unobservable, because their modulation on electron
coherence is negligible~\cite{Szhao2011atomic}. Applying rf control
at the delay window, we can directly detect, identify, and control
type-h and -d pairs.

\medskip{}

\noindent \emph{1. type-h pair} ($|\delta_{j,k}|\gg|d_{j,k}|$)

\medskip{}

For type-h pairs, homonuclear spin-flip processes are suppressed and
therefore the precession frequency of each nuclear spin is split by
$d_{jk}$ by the internuclear interaction $H_{\text{dip}}=d_{jk}I_{j}^{z}I_{k}^{z}$.
The nuclear spin flip by the rf pulse at the delay window is conditional
to the state of the other spin in the nuclear spin pairs. Similar
to the calculation of single spin, when we apply a rf pulse to rotate
the spin $j$, the population left in the original equal superposition
state is
\begin{equation}
P_{\text{NV}}=\frac{1}{2}+\frac{1}{4\mathcal{N}}\text{Tr}(U_{+}U_{-}^{\dagger}+U_{-}U_{+}^{\dagger}),
\end{equation}
with $\mathcal{N}=4$ for the dimension of the spin pair. Here $U_{\pm}=e^{-i\theta I_{j}^{x}\otimes|\uparrow\rangle\langle\uparrow|}e^{\mp i2\eta A_{j}^{\parallel}\tau I_{j}^{z}\otimes|\uparrow\rangle\langle\uparrow|}$
for applying a $\theta_{\text{rf}}=\pi$ spin flip on nuclear spin
$j$ conditioned on the state $|\uparrow\rangle$ of spin $k$. The
corresponding population signal $P_{\text{NV}}=1/2+[1+\cos(2\eta A_{j}^{\parallel}\tau)]/4 \geq 1/2$
for applying $\theta_{\text{rf}}=\pi$ on resonant to the spin $j$.
Note that the signal is different from the case of single spin, as
shown in Figs.~\ref{fig:smFigPair} (a) and (c).

\medskip{}

\noindent \emph{2. type-d pair} ($|\delta_{j,k}|\ll|d_{j,k}|$)

\medskip{}

For type-d pair of homonuclear spins, the interaction takes the form
$H_{\text{dip}}=d_{jk}(3I_{j}^{z}I_{k}^{z}-\boldsymbol{I}_{j}\cdot\boldsymbol{I}_{k})/2$
under a strong magnetic field~\cite{Svandersypen2005nmr}. For the
nuclear spins of $I=1/2$, the composited spin cluster has a singlet
state with a composited spin $J=0$, $|s_{\text{n}}\rangle=(|\uparrow\downarrow\rangle-|\downarrow\uparrow\rangle)\sqrt{2}$.
The triplet states with $J=1$ are $|1_{\text{n}}\rangle=|\uparrow\uparrow\rangle$,
$|0_{\text{n}}\rangle=(|\uparrow\downarrow\rangle+|\downarrow\uparrow\rangle)/\sqrt{2}$,
and $|-1_{\text{n}}\rangle=|\downarrow\downarrow\rangle$. A radio
frequency control $H_{rf}=\gamma_{\text{n}}B_{x}\cos(\omega_{rf}t)(I_{j}^{x}+I_{k}^{x})$
can be written as
\begin{equation}
H_{rf}=\sqrt{2}\gamma_{\text{n}}B_{x}\cos(\omega_{rf}t)(|1_{\text{n}}\rangle+|-1_{\text{n}}\rangle)\langle0_{\text{n}}|+\text{h.c.}.
\end{equation}
Tuning the rf frequency $\omega_{rf}$ around the splitting between
$|0_{\text{n}}\rangle$ and $|\pm1_{\text{n}}\rangle$, the nuclear
spins are rotated. The nuclear state $|0_{\text{n}}\rangle$
has an energy of $-d_{jk}/2$, while the energies for $|\pm1_{\text{n}}\rangle$
are $\pm\omega_{\text{n}}+d_{jk}/4$, where $\omega_{\text{n}}$
is the nuclear precession frequency shifted by the hyperfine field.
Therefore, the transition frequencies between $|0_{\text{n}}\rangle$
and $|\pm1_{\text{n}}\rangle$ are $\omega_{n}\pm 3d_{jk}/4$,
shifted by the dipolar coupling. Different from the case of single
spins that a spin flip requires a rf driving time $2\pi/(\gamma_{\text{n}}B_{x})$,
here the effective rf control field is increased and transitions between
$|0_{\text{n}}\rangle$ and $|\pm1_{\text{n}}\rangle$ can be finished
in a time of $\sqrt{2}\pi/(\gamma_{\text{n}}B_{x})$. This time difference
is a signature to distinguish the signals from that of single spins.
Before a spin flip of the nuclear spin pair, the interaction Hamiltonian
with the NV electronic spin reads $H=\eta\sigma_{z}(A_{j}^{\parallel}I_{j}^{z}+A_{k}^{\parallel}I_{k}^{z})$,
i.e., $H=\eta\sigma_{z}A^{\parallel}(|+1_{\text{n}}\rangle\langle+1_{\text{n}}|-|-1_{\text{n}}\rangle\langle-1_{\text{n}}|)$.
After a time $\tau$, we flip the electron spin and the nuclear triplet
state, e.g., with a transition $|0_{\text{n}}\rangle\leftrightarrow|-1_{\text{n}}\rangle$
in a delay window. We can achieve an effective interaction $H=\eta\sigma_{z}A^{\parallel}(-|0_{\text{n}}\rangle\langle0_{\text{n}}|+|-1\rangle\langle-1|)$
after the delay window. After another delay time $\tau$, the join
evolution up to single qubit operations reads $U=\exp\left[-i\eta\sigma_{z}A_{z}\tau(|+1_{\text{n}}\rangle\langle+1_{\text{n}}|-|0_{\text{n}}\rangle\langle0_{\text{n}}|)\right]$.
The electron spin population modulated by the spin pair signal is
\begin{equation}
P_{\text{NV}}=\frac{1}{2}+\frac{1}{4}[\cos(2\eta A_{z}\tau_{z})+1].
\end{equation}
by using $P_{\text{NV}}=1/2+\text{Tr}(U_{+}U_{-}^{\dagger}+U_{-}U_{+}^{\dagger})/(4\mathcal{N})$
with $\mathcal{N}=4$.

In summary for type-d pairs, the rf pulse drive the transitions between
$|0_{\text{n}}\rangle$ and $|\pm1_{\text{n}}\rangle$ of the nuclear
spin triplet, with a frequency difference of $3d_{jk}/2$ and the
Rabi frequency $\sqrt{2}$ times of the one for single spins, as shown
in Figs.~\ref{fig:smFigPair} (b) and (d). 

Therefore the delayed entanglement echo enables the detection, identification,
and control of differnt types of spin pairs, and it is a useful tool to 
extract information of more complicated spin clusters.

\section{Combining the interaction windows with dynamical decoupling}

We can further protect the NV centre during the interaction windows with
DD. Application of a DD pulse sequence (with the sequence duration $\tau$) 
in an interaction window,
gives the interaction of the form $H_{\text{int}}\approx(\eta/2) f_{k_{\text{DD}}}^{\text{s}}\sum_{j}^{\prime}A_{j}^{\perp}\sigma_{z}I_{j}^{x}$
for symmetric sequences (see Sec.~\ref{sub:DD-Effective-interaction-Hamiltonia}).
Here the summation $\sum_{j}^{\prime}$ is over the spins that are
not resolved by the DD sequence (with a frequency uncertainty of $\sim1/\tau$).
The interaction Hamiltonian takes the same form as the effective Hamiltonian
under a strong magnetic field, but with the spin operators $I_{j}^{x}$
instead of $I_{j}^{z}$ (as well as the coupling constants).

In this manner, the effective interaction $H_{\text{int}}$ during the interaction 
windows does not commute with the spin precession during the delay window. After the 
evolution during the delay window driven by $\omega_{j}I_{j}^{z}$, the nuclear spins 
could have suffered different evolutions on $I_{j}^{x}$ because of possible differences 
of $\sim1/\tau$ in their precession frequencies. This static inhomogeneous of spin
precession can be removed by DD on those nuclear spins. A rf $\pi$
pulse on the nuclear spins effectively reverses the evolution driven by the precession
Hamiltonian $\omega_{j}I_{j}^{z}$. Therefore, the interaction Hamiltonians
in the two interaction windows before and after the delay window are
the same, when we apply a two-pulse CP sequence during the delay window.
With a microwave $\pi$ pulse applied on the NV electron spin before
the second interaction window, the coherence of NV electron spin is
preserved by the delayed entanglement echo at the end of the second
interaction window.

To address desired nuclear spins, in the delay window we apply rf driving with the frequencies
on the target spins to rotate the target nuclear spins. The azimuthal direction $\phi_{\text{rf}}$ 
of rf driving field is controlled by the rf phase and we choose $\phi_{\text{rf}}$ the same as 
the pulse direction of the CP sequence in the rotating frame of nuclear spin precession. When 
the rf driven rotation does not commute with the interaction Hamiltonian, it breaks the erasing 
process of delayed entanglement echo on the target spins, and thereby, we address the nuclear 
spins in a highly selective way. On the other hand, when the rf direction is parallel to the 
azimuthal angle of a target nuclear spin, the rf driving commutes with the interaction window
and the electron-nuclear entanglement is removed after the echo. By measuring the rf phases 
$\phi_{\text{rf}}$ which cause vanishing signal dips, we obtain the relative directions of 
nuclear spins.

\section{Simulation details}

The $^{13}\text{C}$ spins of the diamond samples are randomly distributed
around the NV centre. In simulations for NV dynamics, we randomly
distribute $^{13}\text{C}$ spins around the NV centre and select
samples that do not contain $^{13}\text{C}$ nuclei within a distance
of $0.714$ nm from the NV centre (corresponding to 274 atomic sites),
so that the hyperfine interactions between the $^{13}\text{C}$ nuclei
and NV electron spin are simply described by the dipolar coupling Eq.~(\ref{eq:Aj}).
The probability of getting this kind of samples is $\sim5\%$ for
natural abundance of $1.1\%$ and is higher for lower abundances.
Because of low spin concentration, simulations are accurate enough
by grouping nuclear spins into interacting clusters and neglecting the intercluster
interactions~\cite{Smaze2008electron}. Because of the application of 
control fields {[}magnetic fields of the form $B_{\text{c}}\cos(\omega_{\text{c}}t+\phi_{\text{c}})${]},
the total Hamiltonians for the simulations become time-dependent.
To simulate the control fields, we sample the control fields in a time
step of the minimum values of $0.01\times2\pi/\omega_{\text{c}}$.

We adopt the coordinate system $\hat{z}=[111]/\sqrt{3}$ along the
symmetry axis of NV centre and the orthogonal unit vectors $\hat{x}=[1\bar{1}0]/\sqrt{2}$
and $\hat{y}=[11\bar{2}]/\sqrt{6}$ to record the positions of $^{13}\text{C}$
spins $\boldsymbol{r}_{j}=[\boldsymbol{r}_{j}\cdot\hat{x},\boldsymbol{r}_{j}\cdot\hat{y},\boldsymbol{r}_{j}\cdot\hat{z}]$,
which are measured relative to the location of the NV electron spin
at the origin $[0,0,0]$.

\begin{figure*}
\includegraphics[clip,width=0.9\textwidth]{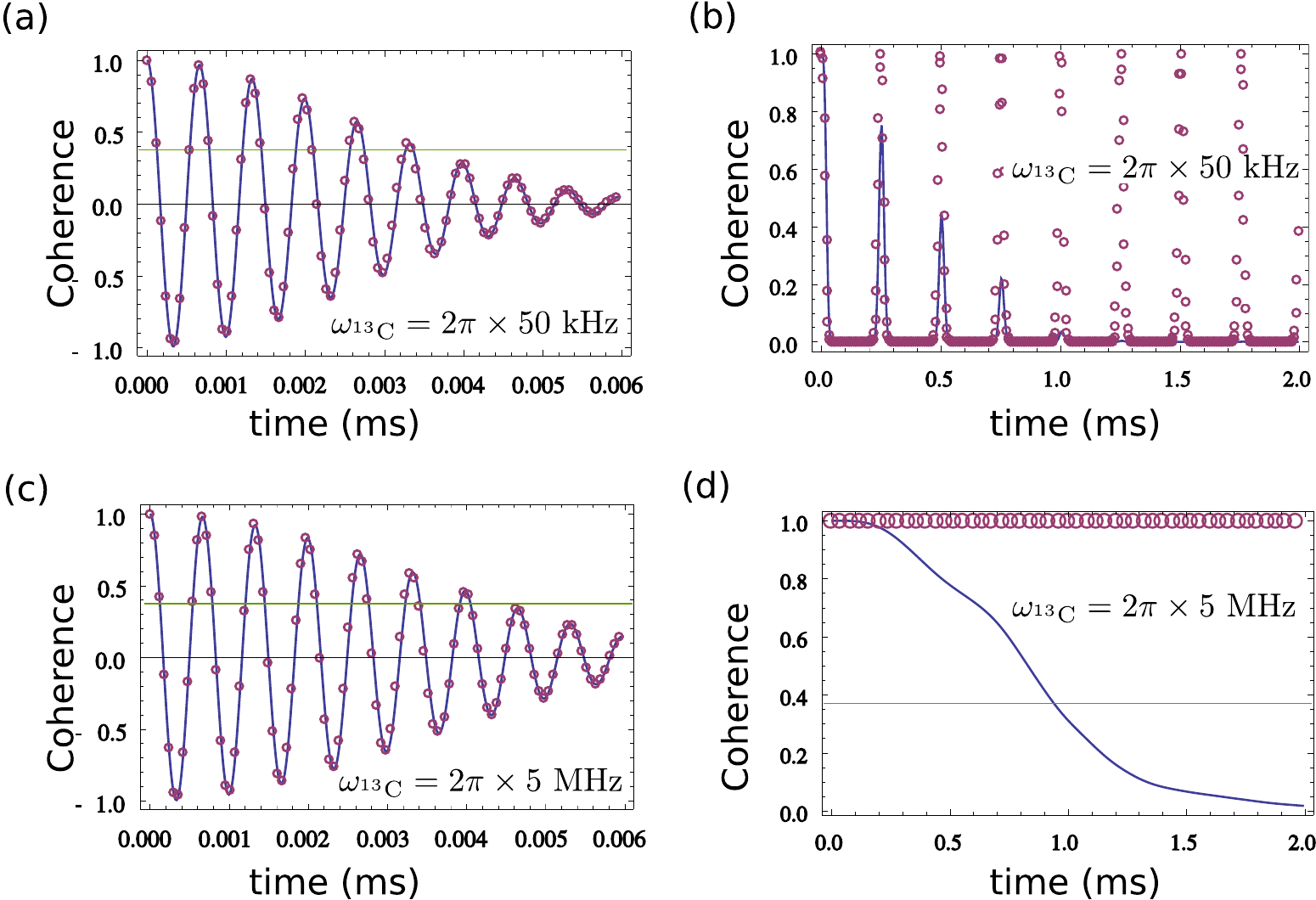}\caption{\label{fig:smFigHahn} \textbf{Coherence of NV electron spin in interacting
and non-interacting spin baths.} (a) Free evolution of the
coherence (blue solid line) of NV electron spin in a spin bath used in Fig.~2 of the
main text. (b) Coherence of NV electron
spin with the application of a spin echo pulse at the middle of the
evolution time (blue solid line). (c) As in (a), but
with a much stronger magnetic field. (d) As in (b),
but with a magnetic field as in (c). In (a)-(d),
the circles corresponds to the case without nuclear-nuclear interactions.
Under strong magnetic fields and without nuclear-nuclear interactions,
the NV electron spin can have a much longer coherence time.}
\end{figure*}

The sample used for Fig.~2 of the main text contains the host nitrogen
and 736 $^{13}\text{C}$ spins. The simulations are converged for
clusters with up to 7 spins and intercluster interactions $\lesssim2\pi\times70$
Hz. For this sample, a spin echo $\pi$ pulse extends the electron
coherence times from $T_{2}^{*}\approx4$ $\mu$s to $\sim1$ ms under
magnetic fields much larger than the hyperfine fields at the nuclear spins
(see Fig.~\ref{fig:smFigHahn}), consistent with experiments~\cite{Scramer2015repeated} 
and theories~\cite{Szhao2012decoherence}. Fig.~\ref{fig:smFigHahn} (d) also shows that the NV coherence time can be much
longer if the nuclear-nuclear interactions are suppressed.

In simulations for the results of Fig.~3a of the main text,
the electron spin relaxation is solved by Lindblad master equations.
The signals comes from the addressing to an isolated nuclear spin
located at $\boldsymbol{r}_{j}=[0.0,-1.9635,-0.8925]$ nm with the
hyperfine field components $A_{j}^{\parallel}=2\pi\times1.49$ kHz
and $A_{j}^{\perp}=2\pi\times2.93$ kHz. The $^{13}\text{C}$ memory
qubit located at $[-0.714,0.0,0.357]$ nm has $A_{m}^{\parallel}=-2\pi\times31.26$
kHz and $A_{m}^{\perp}=2\pi\times29.24$ kHz. We use AXY sequences
to realize the iSWAP gate (see Sec.~\ref{sub:Storage-to-carbon-13})
on the $^{13}\text{C}$ qubit with a gate time $2t_{g}\approx318$
$\mu$s, using a total number of $\sim152$ composite $\pi$ pulses
(explicitly, 760 elementary $\pi$ pulses since one composite pulse
in AXY sequences has 5 elementary $\pi$ pulses).

In Fig.~3b of the main text, the two spins in a C-C bond
are located at $\boldsymbol{r}_{j}=[-1.2495,0.714,-0.1785]$ nm and
$\boldsymbol{r}_{k}=[-1.33875,0.80325,-0.26775]$ nm, which imply
a dipolar coupling of $d_{j,k}=2\pi\times1.37$ kHz. The hyperfine
components $A_{j}^{\parallel}=-2\pi\times4.94$ kHz and $A_{j}^{\perp}=2\pi\times5.33$
kHz for spin $j$, while $A_{k}^{\parallel}=-2\pi\times3.72$ kHz
and $A_{k}^{\perp}=2\pi\times4.2$ kHz for spin $k$.

In Fig.~3c of the main text, the two separated spins located
at $\boldsymbol{r}_{j}=[0.0,-1.9635,-0.8925]$ nm (i.e., the target
spin in Fig.~3a) and $\boldsymbol{r}_{k}=[0.0,1.2495,1.9635]$
nm have similar hyperfine components. The values for the second spin
$A_{k}^{\parallel}=2\pi\times1.43$ kHz and $A_{k}^{\perp}=2\pi\times2.28$
kHz. We protect the interaction window with $\tau=0.5$ ms, by using
CP sequences with 1000 microwave $\pi$ pulses (corresponding to 200 composite
$\pi$ pulses if we use AXY sequences). The two $\pi$ pulses on the
$^{13}\text{C}$ spins are implemented by rf fields with a Rabi frequency
of $2\pi\times20$ kHz.

In the simulation with optical illumination used in Fig.~3d
of the main text, we adopt the Lindblad model and parameters of the
experimental paper~\cite{Smaurer2012room}. The memory $^{13}\text{C}$
spin is similar to the one used in ref.~\onlinecite{maurer2012room},
with the location $\boldsymbol{r}_{m}=[-0.108,-0.295,-1.74]$ nm and
hyperfine components $A_{m}^{\parallel}=-2\pi\times1.69$ kHz and
$A_{m}^{\perp}=2\pi\times5.4$ kHz. We use spin lock technique on
the NV electron qubit with the spin lock frequency on resonant to
the precession frequency of the $^{13}\text{C}$ memory to perform
a iSWAP gate (see Sec.~\ref{sub:Storage-to-carbon-13}). To implement
a complete SWAP gate, we use delayed entanglement echo on the $^{13}\text{C}$
memory after the iSWAP operation. The delay window for the $^{13}\text{C}$
memory uses a 20-pulse CP sequence with duration $\approx100$ $\mu$s
for a protected rf $\pi$ gate. The total SWAP gate time for the simulation
is $\approx584$ $\mu$s, which can be reduced by a factor of two
if we use the electron levels $m_{s}=\pm1$. The gate can be further protected by storing the electron
state to the nitrogen spin when applying the delay window for the $^{13}\text{C}$
memory. Using a $^{13}\text{C}$ memory more strongly coupled to the
NV electron can also significantly reduce the required SWAP gate times.
The proton spin for detection is located $4$ nm away from the NV
centre, with $\boldsymbol{r}_{j}=[2.31,2.31,2.31]$ nm (hence $A_j^{\perp}=0$ and the spin is 
hard to detect by traditional DD).

The procedure to detect the chemical shifts of proton spins with optical
illumination is the following. We first pump the NV electron spin
by optical field to initialize the NV electron spin to the state $|0\rangle$
with a fidelity $82\%$ (the fidelity is obtained for the parameters in ref.~\onlinecite{Smaurer2012room} 
and it is higher for better samples), which is followed by using a swap operation
to polarize the $^{13}\text{C}$ memory spin. Then we use optical
pumping again and a microwave pulse to initialize the NV electron
to a superposition state $(|0\rangle+|1\rangle)/\sqrt{2}$. After
the initialization of NV electron spin and memory qubit, we let the
whole system freely evolve for a time of $\tau$ to generate electron-nuclear
entanglement (which can be protected by DD as shown in the main text).
Then we store the NV electron spin to the memory spin by a swap gate.
Subsequently we use optical illumination to decouple electron-nuclear
coupling for applying a rf pulse with the length $t_{\text{rf}}\approx80$
ms. The carry frequency of rf driving is set to the target proton
spins. After optical illumination, we wait for 2 $\mu$s to relax
the NV electron spin back to $|0\rangle$ state. Subsequently, we
use a swap gate to retrieve the quantum state of NV electron spin
and to re-popularize the ancillary $^{13}\text{C}$ spin. Finally,
we apply a microwave $\pi$ pulse on the electron spin and wait for
another interaction window of time $\tau$ before readout of the electron
spin state.  We can increase the interaction to target spins by using 
the NV levels $m_{s}=\pm1$, as described in the main text.


\end{document}